\documentclass[prb,aps,twocolumn]{revtex4}
\usepackage{graphicx}
\usepackage{amssymb}

\begin{document}
\title{Kondo effect in coupled quantum dots: a Non-crossing approximation study.}
\author{Ram\'on Aguado$^{1,2}$} 
\email{raguado@icmm.csic.es}
\author{David C. Langreth$^{1}$}
\affiliation{1-Center for Materials Theory, Department of Physics and Astronomy, Rutgers University, Piscataway, NJ
08854-8019, USA.}
\affiliation{2-Departamento de Teor\'{\i}a de la Materia Condensada, Instituto de Ciencia de Materiales de Madrid, 
CSIC, Cantoblanco 28049, Madrid, Spain.}

\date{\today}

\begin{abstract}
The out-of-equilibrium transport properties of a double quantum
dot system in the Kondo regime are studied theoretically by means
of a two-impurity Anderson Hamiltonian with inter-impurity hopping.
The Hamiltonian, formulated in slave-boson language, is solved by means of a 
generalization of the non-crossing approximation (NCA) to the present problem.
We provide benchmark calculations of the predictions of the NCA
for the linear and nonlinear transport properties of coupled quantum dots in the Kondo regime.
We give a series of predictions that can be observed experimentally
in linear and nonlinear transport measurements through coupled quantum dots.
Importantly, it is demonstrated that measurements
of the differential conductance ${\cal G}=dI/dV$,
for the appropriate values of voltages and inter-dot tunneling couplings, can give a direct observation
of the coherent superposition between the many-body Kondo states of
each dot. This coherence can be also detected in the linear transport through the 
system: the curve linear conductance vs temperature is non-monotonic, with a maximum at a 
temperature $T^*$ characterizing quantum coherence between both Kondo states. 
\end{abstract}
\pacs{PACS numbers: }
\maketitle

\section{INTRODUCTION}
The recent observations of Kondo physics in the electronic transport properties of 
quantum dots (QD's)  \cite{Gold,Leo,Stutt,Blick,Wilfred}, a tiny semiconductor box containing
a few interacting electrons\cite{dot},
have opened new promising directions for experimental and theoretical research of this phenomenon,
one of the paradigms in condensed matter physics.\\
The Kondo effect appears in dilute alloys containing localized moments as a crossover from weak
to strong coupling between itinerant electrons of the host non-magnetic
metal and the unpaired localized electron of the magnetic impurity as the
temperature ($T$) is reduced well below the Kondo temperature ($T_K$)
.\cite{Hew}
Due to spin exchange interaction, a many-body spin singlet state is formed
between the unpaired localized electron and the itinerant electrons with energies close to the Fermi energy of the metal. 
This singlet is reflected
in the density of states (DOS) of the impurity as a narrow peak at low frequencies: the Abrikosov-Suhl (AS) or Kondo resonance.
This many-body resonance in the DOS is responsible for anomalous properties:
In the intermediate temperature regime $T\gtrsim T_K$, this effect leads to logarithmic corrections 
to the magnetic susceptibility $\chi(T)$, the linear specific heat coefficient $\gamma(T)$ 
and the resistivity $\rho(T)$.
Below the Kondo temperature, it leads to saturated
behavior of the magnetic susceptibility $\chi(T)=const$, the linear specific heat coefficient $\gamma(T)=const$
and the resistivity $\rho(T)-\rho(0)\sim T^2$ (Fermi liquid behavior). 
The Kondo effect, being one of the most widely studied phenomenon in condensed matter physics, has been
studied for decades.
The first manifestations of the Kondo effect in a linear transport property, 
namely a resistance minimum at finite temperatures, date back to the early 30's.\cite{Hew}
Zero-bias anomalies in the nonlinear tunneling conductance, the hallmark of Kondo physics, 
were first observed during the 60's. \cite{Logan} 
Finally, the first measurements of nonlinear transport through a single Kondo impurity 
were reported in the mid 90's.\cite{Ralph}\\
In recent years, spectacular advances in nanotechnology have made it possible to
experimentally study Kondo physics in quantum dots.\cite{Gold,Leo,Stutt,Blick,Wilfred}
These truly impressive experiments confirm early theoretical predictions that transport 
through quantum dots in the Coulomb blockade regime should exhibit
Kondo physics at low enough temperatures.\cite{Kon}
QD's provide the intriguing opportunity to control and modify the Kondo effect experimentally:
the continuous tuning of the relevant parameters
governing the Kondo effect \cite{Hew} as well as the possibility of studying Kondo physics
when the system is driven out of equilibrium, either by dc \cite{Hers1,Win1,Levy,NCAneq2,NCAneq2b,NCAneq2c,magnetization}
or ac voltages \cite{Schoe,Ng,Hers2,Rosa,Avis,Kam,Nord}, pave the way for the study of 
strongly correlated electron physics in artificial systems. Moreover, they  provide a 
unique testing ground in which to investigate the interplay of strongly correlated electron physics, 
quantum coherence and non-equilibrium physics.
More sophisticated configurations of QD's in the Kondo regime constitute
a growing area of intense investigations, both from the theoretical and experimental sides.
Time dependent Kondo physics \cite {time1,time2,time3,time4}, Kondo physics in integer-spin QD's \cite{Integ} or
QD's embedded in Aharonov-Bohm rings \cite{Ring} 
are examples of such configurations.                          
\par
The study of Kondo physics in mesoscopic or nanoscopic systems is not limited to QD's. 
We can mention here the recent observation of Kondo physics in single atoms \cite{atom}, 
molecules \cite{molecule}, carbon nanotubes \cite{nanotub}, scanning tunneling
microscopy (STM) experiments of magnetic impurities in quantum corrals \cite{corral} or
the anomalous energy relaxation in voltage-biased quantum wires and its relation to two-channel Kondo
physics.\cite{twochan}
\par
In this paper we will focus on another configuration: a system of two coupled quantum dots in the Kondo regime.
In view of the  recent experimental advances in the study of
quantum coherence in coupled quantum dots (DQD) \cite{Oos,Tosh,Blick2} and the aforementioned studies of Kondo physics in quantum dots,
it is a timely question to ask what happens when a system consisting of two quantum dots in the Kondo 
regime, coupled to each other by means of a tunneling barrier,
is driven out-of-equilibrium, and how the interplay
of strongly correlated electron physics,
quantum coherence and non-equilibrium physics leads to new physical scenarios.
Previous theoretical studies of this problem at equilibrium
have focused on aspects of quantum coherence in this system \cite{Aon,Aon2} 
and on the competition between Kondo effect and anti-ferromagnetic coupling generated
via exchange \cite{Geor,Num,Aon2} or via capacitive coupling between dots \cite{Andr}.
There have hitherto been only few attempts to attack this problem in a non-equilibrium situation by means
of different techniques: the equation-of-motion technique (EOM) \cite{Ivan}, the so-called resonant tunneling approximation \cite{Poj} 
(valid for $T>T_K$ and equivalent to the EOM method of Ref. \onlinecite{Win1})
and slave-boson mean field theory \cite{Agua1}. 
Here, we present an approach which, for the first time, tackles with this non-equilibrium problem in a 
non-perturbative, fully self-consistent and conserving way. Our approach is based on a generalization
of the so-called "non-crossing approximation" (NCA) \cite{NCA1,NCA2} to the present problem.
\par
The system of two coupled QD's can be modeled by means of two Anderson impurities, 
each of them coupled to a different Fermi sea, and coupled together
by means of an inter-impurity hopping term. 
Quantum impurity models such as the Kondo and the Anderson impurity problems were first 
introduced in the 60's trying to explain the aforementioned anomalous properties of metals in the presence of magnetic impurities.
More generally, this class of problems and their generalization to the lattice
constitute one of the paradigms of modern condensed matter theory.
They typically consist of conduction electrons coupled to sites where there is a strong on-site 
Coulomb interaction, and are believed to describe the rich physics of many different strongly 
correlated electron systems like, for example, the heavy 
fermion compounds. \cite{Heavy}
Examples of these models are the Hubbard model, the $t-J$ model and the 
Anderson or Kondo lattice models.
In all these models the main difficulty resides in the fact that usual perturbation theory does not apply.
On one hand, if the on-site Coulomb repulsion exceeds the band width, conventional many-body perturbation theory
in the on-site repulsion does not work. On the other hand, the obvious alternative of perturbation theory in the kinetic energy 
is not valid due to the non-canonical commutation relations of the field operators
in the atomic limit. At the heart of the problem is the characteristic feature of strongly correlated electrons: the dynamics is
constrained to a subspace of the total Hilbert space. 
For instance, in the atomic limit of the Hubbard model each lattice site can either be empty $|0\rangle$, singly occupied $|\uparrow\rangle$, 
$|\downarrow\rangle$
or doubly occupied $|\uparrow\downarrow\rangle$. 
The operators describing these states and the transitions among them, the Hubbard operators, are neither fermions nor bosons which 
precludes the application of usual perturbation theory (Wick's theorem does not apply). One way of circumventing this difficulty
is the auxiliary particle representation pioneered by Abrikosov, who first represented local spins by pseudo-fermions \cite{slave1}, 
and later by Barnes\cite{slave2} and Coleman \cite{slave3}
and consists of 
describing
each of the states (for each site) as created out of the vacuum from the application of a creation operator 
(bosonic for $|0\rangle$ and  $|\uparrow\downarrow\rangle$ and fermionic for $|\uparrow\rangle$ and $|\downarrow\rangle$ due to 
quantum statistics).
Each site has to be in one of the four states, this is accomplished by constraining the number of auxiliary particles to one. 
Slave particle representations allow one to work with usual quantum field theory methods
provided one works in the constrained subspace of the Hilbert space where the number of auxiliary particles 
is one.
In particular
in the limit of infinite on-site interaction, which is case we shall consider in the following, each site can be described 
by one boson $|0\rangle\equiv b^\dagger|vac\rangle$ and two fermions $|\uparrow\rangle\equiv f^\dagger_\uparrow|vac\rangle$, 
$|\downarrow\rangle\equiv f^\dagger_\downarrow|vac\rangle$. This particular version of the auxiliary particle representation has 
been termed slave boson (SB) representation after Coleman.\cite{slave3} 
\par
Within the SB formulation two non-perturbative approaches can be applied to N-fold degenerate Anderson-impurity models:\\
{\it i)} The mean-field approximation (MFA) of the slave boson field,\cite{slave3,Read} 
only valid for describing spin fluctuations in the Kondo regime, 
correctly generates the low energy scale $T_K$ and leads to local Fermi-liquid behavior at zero-temperature. 
The MFA, however, does suffer from two drawbacks: a) it leads always to local Fermi liquid behavior, 
even for multichannel models; b) The MFA has a phase transition 
(originating from the breakdown of the local gauge symmetry of the problem) 
that separates the low temperature state from the high temperature
local moment regime. This later problem may be corrected by including 1/N fluctuations around the mean-field solution.\cite{transi}
The generalization of the SBMFA to the present problem, 
two coupled quantum dots in a non-equilibrium situation has been studied in Ref. \onlinecite {Agua1}\\ 
{\it ii)} The Non-crossing approximation (NCA) \cite{NCA1,NCA2}
is the lowest order self-consistent, fully conserving and $\Phi$ derivable theory in the Baym sense.\cite{Baym}
It is well known that the NCA fails in describing the low-energy Fermi-liquid regime. 
Neglect of vertex corrections prevents from a proper description of low-energy excitations. Nevertheless, the NCA
has proven to give reliable results
for temperatures down to a fraction of $T_K$.\cite{Costi}  
The NCA gives better results in multichannel cases, where the correct 
non-Fermi liquid behavior is obtained \cite{cox}.
Nonetheless, Kroha {\it et al} \cite{Kroha2} have shown in a series of papers that it is possible to develop systematic corrections
to the NCA's $\Phi$ functional that cure the low-temperature pathologies of the NCA. 
These systematic corrections (the so-called
"conserving T-matrix approximation") are able to describe Fermi liquid and Non-Fermi liquid regimes
on the same footing. It is also possible to formulate NCA equations for finite $U$ by symmetrizing the usual NCA diagrams with respect to empty and doubly occupied local states.\cite{Kroha3} 
\par
The generalization of the NCA to time-dependent phenomena
 was developed by Langreth and collaborators in a 
series of papers \cite{NCAneq1,NCAneq1b} (see also Ref. \onlinecite{NCAneq1c}).
and later applied to non-equilibrium transport through quantum dots \cite{Win1,NCAneq2,Inos,NCAneq3,NCAneq4} 
and other 
mesoscopic systems.\cite{NCAneq5}\\
In this work, the NCA is generalized to cope with the present problem, 
namely two Anderson impurities, coupled to each other by a 
tunneling barrier,
which are in a non-equilibrium situation.
\par
The paper is divided as follows: 
\begin{figure}
\includegraphics [width=0.5\textwidth,clip] {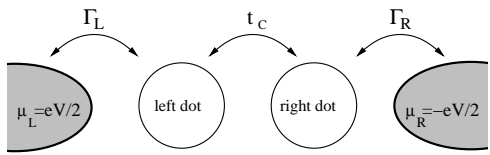}
\caption{Schematic diagram of the double dot system studied in this paper. 
Each dot is coupled independently to one lead with couplings 
$\Gamma_L$ and $\Gamma_R$ respectively, $t_C$ is the interdot tunneling term.
Note that
the role of the inter-dot term is twofold: firstly, it generates quantum coherence between the 
two quantum dots; 
secondly, it establishes a non-equilibrium situation,
when the chemical potentials are different there is a bias voltage across the system $\mu_L-\mu_R=eV$ 
and, then, a flow of electrical current through the double dot system.}  
\label{fig1}
\end{figure}  
In Sec. II we formulate the Hamiltonian (the general form and its slave-boson formulation) which describes 
the problem.
In Sec. III we briefly review the non-equilibrium Green's function technique, real time Dyson equations
for the retarded and lesser Green's function, that we use in order to formulate the problem in its 
fully non-equilibrium form.
In Sec. IV we present our generalization of the NCA technique to the problem: 
In Sec. IVA the self-energies obtained within our scheme are presented and discussed.
In Sec. IVB we derive the physical two-particle correlation functions within the NCA approach.
In Sec. IVC we present the fermion and boson selfenergies after the projection onto the restricted Hilbert space.
We present and discuss in Sec. V various model calculations for the density of states (DOS), 
both in equilibrium and 
non-equilibrium situations, linear conductance, non-linear current and
non-linear differential conductance. 
We give a series of predictions for the current and finite voltage 
differential conductance
which are relevant for experiments. It is demonstrated that measurements
of the differential conductance ${\cal G}=dI/dV$,
for the appropriate values of voltages and inter-dot tunneling couplings, can give a direct observation
of the coherent superposition between the many-body Kondo states of
each dot. We also give predictions for the temperature dependence of the linear 
conductance and for the 
nonlinear differential conductance in the high-voltages regime, where negative differential conductance 
is obtained for low temperatures and large inter-dot couplings.
An appendix is included to discuss the projection procedure used to deal with the constraint in the Hilbert space.
\section{MODEL}
\subsection{General formulation}
As we mentioned already, the double quantum dot can be modelled as a 
two-impurity Anderson Hamiltonian with an extra term accounting for inter-impurity hopping.  
Each impurity is connected to a different Fermi sea with chemical potential
$\mu_L=\frac{eV}{2}$ and $\mu_R=-\frac{eV}{2}$ respectively ($\epsilon_F=0$).
\begin{widetext}
\begin{eqnarray}
H&=&\sum _{ k_{\alpha\in \{ L,R \}}, \sigma} \epsilon _{k_\alpha} c^\dagger_{k_\alpha,\sigma}
c_{k_\alpha,\sigma} 
+ \sum _{\alpha\in \{ L,R \},\sigma} \epsilon _{\alpha\sigma} d^\dagger_{\alpha\sigma}
d_{\alpha\sigma} 
+V_0\sum _{k_{\alpha\in \{ L,R \}},\sigma }(c^\dagger_{k_\alpha, \sigma} d_{\alpha\sigma}
+ d^\dagger_{\alpha\sigma} c_{k_\alpha, \sigma})
+V_C\sum _\sigma(d^\dagger_{L\sigma}d_{R\sigma}+d^\dagger_{R\sigma}d_{L\sigma})\nonumber\\
&+&U_L n_{L,\uparrow} n_{L,\downarrow}
+U_R n_{R,\uparrow} n_{R,\downarrow}.
\end{eqnarray}
\end{widetext}
The first two terms in the Hamiltonian represent the electrons in the leads and in the dots respectively.
In these hamiltonians, $c^\dagger_{k_{L/R},\sigma}$ ($c_{k_{L/R},\sigma}$) creates (annihilates) an
electron with momentum $k_{L/R}$ and spin $\sigma$ in the left/right lead, and
$d^\dagger_{L/R\sigma}$ ($d_{L/R\sigma}$) creates (annihilates) an electron with spin $\sigma$ in the left/right dot.
$\epsilon_{k_{L/R}}=\epsilon_{k}+\mu_{L/R}=\epsilon_{k}\pm\frac{eV}{2}$ 
and $\epsilon _{\alpha\sigma}$ are the energies in the leads and the dots, respectively.
The third term describes the coupling between each dot and its corresponding lead, and 
determines the coupling strength
$\Delta^{L,R}=\pi V_0^2\sum_{k_{\alpha\in \{ L,R \}}} \delta(\epsilon-\epsilon_{k_{\alpha}})$
(we neglect the k dependency
of the tunneling matrix element for simplicity).
The fourth term describes 
inter-dot tunneling.
In the absence of inter-dot tunneling, this Hamiltonian describes two independent Anderson impurities
each of them coupled to different Fermi seas (typically at different chemical potentials). Note that
the role of the inter-dot term is twofold: firstly, it generates quantum coherence between the impurities; secondly,
it establishes a non-equilibrium situation, 
when the chemical potentials are different there is a bias voltage across the system and, then, 
there is an electrical current flowing through the double dot system.
The last terms describe the on-site electron-electron interaction on each dot where 
$n_{L/R,\sigma}= d^\dagger_{L/R\sigma} d_{L/R\sigma}$ are the number operators for spin $\sigma$ on each dot.
The on-site interaction parameters are $U_L=e^2/2C_L$ and $U_R=e^2/2C_R$ where
$C_{L/R}$ are the dot capacitances.
The neglect of an interdot electron-electron interaction ($\sim U_{\rm interdot} n_L n_R$) with
$U_{\rm interdot}\sim\frac{C_{LR}}{C_L C_R}$
corresponds to the experimentally
accessible limit of small interdot capacitance ($C_{LR}$) as compared with
the capacitances of each QD to the gates, and implies
a vanishing interdot antiferromagnetic coupling from this source.
\cite{interdot1}
Experimentally, these parameters governing the hamiltonian: tunneling couplings, on-site interactions, etc; can
be purposefully modified by external gate voltages \cite{dot} which allows
to study a variety of rich physical phenomena (spin and charge fluctuations regime, non-equilibrium phenomena, etc) 
on the same sample.\cite{Gold,Leo,Stutt,Blick,Wilfred}
\subsection{Slave-particle representation}
We assume $U_L,U_R\rightarrow\infty$ \cite{interdot2}, forbidding double occupancy on each dot. This is a good
approximation for ultrasmall quantum dots in which the on-site interaction 
is much larger
than the coupling strength $\Delta^{L,R}$ (typically more than one order of magnitude).
In the limit of $U_L,U_R\rightarrow\infty$ (i.e, $C_L,C_R\rightarrow 0$) 
we can write the Hamiltonian
(1) in terms of auxiliary pseudo-fermions and slave boson operators plus constraints:
\begin{widetext}
\begin{eqnarray}
H&=&\sum _{ k_{\alpha\in \{ L,R \}}, \sigma} \epsilon _{k_\alpha} c^\dagger_{k_\alpha,\sigma}
c_{k_\alpha,\sigma}
+ \sum _{\alpha\in \{ L,R \},\sigma} \epsilon _{\alpha\sigma} f^\dagger_{\alpha\sigma}
f_{\alpha\sigma}
+\frac{t_C}{N}\sum _\sigma(f^\dagger_{L\sigma}b_L b^\dagger_R f_{R\sigma}
+f^\dagger_{R\sigma}b_R b^\dagger_L f_{L\sigma})\nonumber\\
&+&\frac{V}{\sqrt N}\sum _{k_{\alpha\in \{ L,R \}},\sigma }(c^\dagger_{k_\alpha, \sigma} b^\dagger_\alpha 
f_{\alpha\sigma}
+ f^\dagger_{\alpha\sigma} b_\alpha c_{k_\alpha, \sigma})
\end{eqnarray}
\end{widetext}
In the slave boson representation, the annihilation operator for electrons in the QD's,
$d_{\alpha\sigma}$ is decomposed into the SB operator $b^\dagger_\alpha$
which creates an empty state and a pseudo fermion operator $f_{\alpha\sigma}$ which annihilates
the singly occupied state with spin $\sigma$ in the dot $\alpha$:
$d_{\alpha\sigma}\rightarrow b^\dagger_\alpha f_{\alpha\sigma}$
($d^\dagger_{\alpha\sigma}\rightarrow f^\dagger_{\alpha\sigma}b_\alpha$).
Note that we have re-scaled the hopping parameters $V_0=\frac{V}{\sqrt N}$ and $V_C=\frac{t_C}{N}$ in order to have a well
defined $1/N$ expansion (N being the degeneracy of each dot).
\begin{figure}
\includegraphics [width=0.45\textwidth,clip] {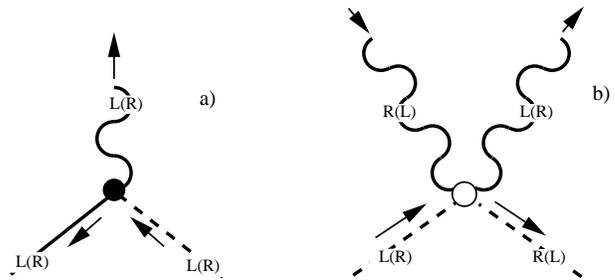}
\caption{Interaction vertices.
Solid, dashed and wavy lines represent lead electron,
pseudo-fermion and slave boson lines, respectively.
Each line carries a left (right) index.
a) Lead-dot hopping vertex $V/\sqrt{N}$ (Full circle). Tunneling of an electron
from the left (right) dot to the left (right) lead is represented as the decay of the
left (right) pseudo-fermion into a left (right) slave boson and left (right)
lead electron.
b) dot-dot hopping vertex $\frac{t_C}{N}$ (Open circle).
Tunneling of an electron
from the left (right) dot to the right (left) dot is represented as the combination of the
left (right) pseudo-fermion with the right (left) slave boson to decay
into a left (right) slave boson and right (left) pseudo-fermion.
Note that this vertex exchanges left and right indexes.}
\label{fig2}
\end{figure}       
Finally, the physical constraint is that we must work in a subspace of the Hilbert space
where the number of auxiliary particles (on each dot) is one, namely:
\begin{eqnarray}
\hat{Q}_L&=&\sum_\sigma f^\dagger_{L\sigma}f_{L\sigma}+b^\dagger_L b_L=1,\nonumber\\
\hat{Q}_R&=&\sum_\sigma f^\dagger_{R\sigma}f_{R\sigma}+b^\dagger_R b_R=1.
\end{eqnarray}
As we mentioned before, these two constraints come from the physical condition that
each dot has to be in one of the three states $|0\rangle$,  $|\uparrow\rangle$ or $|\downarrow\rangle$.
To simplify the notation we consider
henceforth that $\epsilon _{L\sigma}=\epsilon _{R\sigma}=\epsilon _{0}$.
The Hamiltonian (2) has two different kind of fermion-boson interactions 
which are given by the vertices in Fig. 2.
\section{Green's functions and self-energies}
At this point, we have reduced the original problem described by the Hamiltonian in Eq. (1) to a problem
of fermions and bosons interacting through the vertices of  Fig. 2 and subject to the constraints in Eq. (3).
Properties of the physical electrons can be build up from the Green's functions of the pseudo-fermions and slave bosons
(see section IVB).
These Green's functions for the auxiliary
fermions and bosons constitute the basic  building blocks of the theory.
Furthermore, our aim is to study  the out-of-equilibrium properties of the system;
we need, then, a fully non-equilibrium description of the dynamics of the Green's functions of these auxiliary particles.
The appropriate starting point is to derive equations-of-motion (EOM) for the time-ordered double-time Green's 
function of the auxiliary fermion (G) 
and boson (B) fields on a complex contour. 
\begin{figure}
\includegraphics [width=0.45\textwidth,clip]{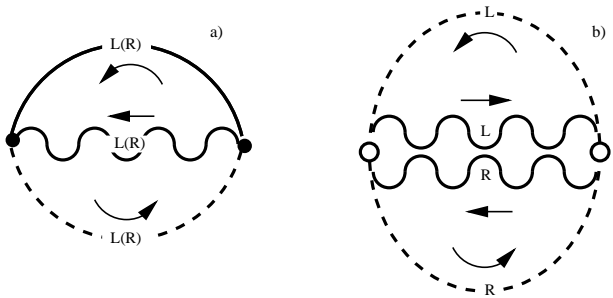}
\caption{Diagrammatic representation of the generating functional $\Phi=\Phi_1+\Phi_2$ of our NCA approximation.
Solid, dashed and wavy lines represent lead electron,
pseudo-fermion and slave boson lines, respectively.
Each line carries a left (right) index.
Full circle: Lead-dot hopping vertex $V/\sqrt{N}$, Open circle:  dot-dot hopping vertex $\frac{t_C}{N}$.
The self-energies are obtained by taking the functional derivative of $\Phi$ with respect
to the corresponding Green's function.
a) Lead-dot functional $\Phi_1$ (leading order $\mathcal{O}(1)$).
b) dot-dot functional $\Phi_2$ (leading order $\mathcal{O}(1/N)$).}
\label{fig3}
\end{figure}     
A rigorous and well established way to derive these EOM was first introduced
by Kadanoff and Baym, \cite{Kadan} and has been related to other non-equilibrium methods 
(like the Keldysh method) by Langreth, see Ref. \onlinecite{Lang} for a review. 
\par
The time-ordered double-time Green's function are defined as (sub-indexes are omitted here):
\begin{eqnarray}
iG(t,t')&\equiv&\langle T_c f(t)f^\dagger(t')\rangle\nonumber\\
iB(t,t')&\equiv&\langle T_c b(t)b^\dagger(t')\rangle.
\end{eqnarray}
Here the time ordering operator $T_c$ and the step functions $\theta$ operate along a contour {\it c} in the complex plane.
It will not matter in the derivation given here whether {\it c} is taken to be the Keldysh contour, the Kadanoff-Baym contour,
or a more general choice.
\par
The time-ordered Green's functions functions can be decomposed in terms of their analytic pieces:
\begin{eqnarray}
iG(t,t')&=&G^{>}(t,t')\theta(t-t')-G^{<}(t,t')\theta(t'-t)\nonumber\\
iB(t,t')&=&B^{>}(t,t')\theta(t-t')+B^{<}(t,t')\theta(t'-t);
\end{eqnarray}
where $G^{<}(t,t')\equiv\langle f^\dagger(t')f(t)\rangle$ 
and $B^{<}(t,t')\equiv\langle b^\dagger(t')b(t)\rangle$ 
are the so-called lesser  Green's functions, 
and $G^{>}(t,t')\equiv\langle f(t)f^\dagger(t')\rangle$ 
and $B^{>}(t,t')\equiv\langle b(t)b^\dagger(t')\rangle$
are the greater ones.\\ 
The retarded propagators can be written in terms of these analytic pieces as:
\begin{eqnarray}
iG^{r}(t,t')&=&[G^{>}(t,t')+G^{<}(t,t')]\theta(t-t')\nonumber\\
iB^{r}(t,t')&=&[B^{>}(t,t')-B^{<}(t,t')]\theta(t-t')\nonumber\\
\end{eqnarray}
The advanced ones can be obtained from $G^{r}(t,t')=[G^{a}(t',t)]^*$.
\par
\begin{widetext}  
The basic starting equations follow directly from the Dyson equations in complex time space:
\begin{eqnarray}
(i\frac{\partial}{\partial t}-\epsilon _{0})G(t,t')&=&\delta(t-t')
+\int_c dt_1 \Sigma(t,t_1)G(t_1,t'),\nonumber\\
i\frac{\partial}{\partial t}B(t,t')&=&\delta(t-t')
+\int_c dt_1 \Pi(t,t_1)B(t_1,t').
\end{eqnarray}
Applying analytic continuation rules \cite{Lang} we can write Dyson equations in real time space which relate 
the lesser and the greater Green's functions with the retarded and advanced ones:
\begin{eqnarray}
(i\frac{\partial}{\partial t}-\epsilon _{0})G^{\gtrless}(t,t')&=&
\int_{-\infty}^{\infty} dt_1 [\Sigma^{r}(t,t_1)G^{\gtrless}(t_1,t')
+\Sigma^{\gtrless}(t,t_1)G^{a}(t_1,t')],\nonumber\\
i\frac{\partial}{\partial t}B^{\gtrless}(t,t')&=&
\int_{-\infty}^{\infty} dt_1 [\Pi^{r}(t,t_1)B^{\gtrless}(t_1,t')
+\Pi^{\gtrless}(t,t_1)B^{a}(t_1,t')]
\end{eqnarray}
The retarded (and advanced) Green's functions follow usual Dyson equations:
\begin{eqnarray}
(i\frac{\partial}{\partial t}-\epsilon _{0})G^{r}(t,t')&=&\delta(t-t')
+\int_{-\infty}^{\infty} dt_1 \Sigma^{r}(t,t_1)G^{r}(t_1,t'),\nonumber\\
i\frac{\partial}{\partial t}B^{r}(t,t')&=&\delta(t-t')
+\int_{-\infty}^{\infty} dt_1 \Pi^{r}(t,t_1)B^{r}(t_1,t').
\end{eqnarray}
The set of Dyson equations is closed by choosing a suitable
approximation for the self-energies $\Sigma$ and $\Pi$, and hence for their analytic pieces
$\Sigma^<$, $\Sigma^>$, $\Pi^<$ and $\Pi^>$. We describe in the following section the non-crossing approximation
used to solve our problem. 
\end{widetext}
\section{Non-crossing approximation (NCA)}
\subsection{Self-energies}
We use the NCA technique \cite{NCA1,NCA2} for obtaining the self-energies 
$\Sigma_{L(R) , \sigma}(t,t')$, $\Pi_{L(R)}(t,t')$
in Eq. (7) and their real time analytic continuations. 
Hereafter, we focus on static non-equilibrium, dc voltages,
the time-translational invariance is thus not broken, i.e  all quantities depend only on 
the time difference $t-t'$. \cite{broken,NDC}
Nonetheless, our NCA equations for the self-energies, see Eqs. (10-11) below, together with Dyson 
equations in real time Eqs. (8-9) are valid for general situations with broken time-translational symmetry by 
just substituting $(t-t')\rightarrow (t,t')$ and solving the fully time-dependent problem.
\par
The generalization of the NCA for time-dependent phenomena was developed by Langreth {\it et al} \cite{NCAneq1,NCAneq1b} and
has been successfully applied to non-equilibrium transport through quantum dots \cite{Win1,NCAneq2,Inos,NCAneq3,NCAneq4},
tunnel junctions and point contacts \cite{NCAneq5}, non-equilibrium dynamics at surfaces and STM studies 
\cite{Plihal1,Plihal2,Plihal3}.
Also, this technique has recently been applied to the study of non-equilibrium dynamics in quantum dots in the Kondo regime \cite{NCAdin}
and to the study of non-equilibrium-induced decoherence \cite{Kroha1} also in quantum dots in the Kondo regime.
\par
As we already mentioned, this technique can be justified as an $1/N$ expansion, at lowest order in perturbation theory,
although it is better regarded as a fully conserving, self-consistent,
and $\Phi$ derivable theory in the Baym sense. \cite{Baym}
\begin{figure}
\includegraphics [width=0.45\textwidth,clip]{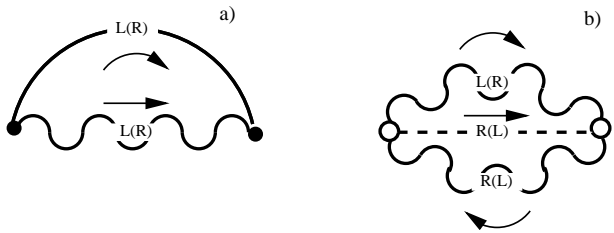}
\caption{Fermion Self-energy. Solid, dashed and wavy lines represent lead electron,
pseudo-fermion and slave boson lines, respectively.
Each line carries a left (right) index. 
Full circle: Lead-dot hopping vertex $V/\sqrt{N}$, Open circle:  dot-dot hopping vertex $\frac{t_C}{N}$.
The leading order of this selfenergy is $\mathcal{O}(1/N)+\mathcal{O}(1/N^2)$}
\label{fig4}
\end{figure}
\begin{figure}
\includegraphics [width=0.45\textwidth,clip]{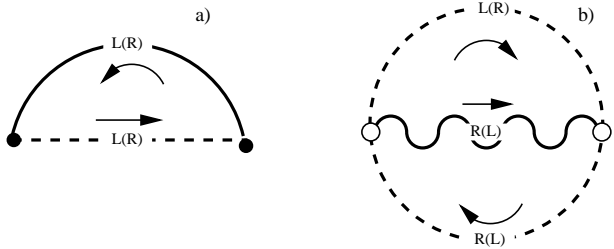}
\caption{Boson Self-energy. Solid, dashed and wavy lines represent lead electron,
pseudo-fermion and slave boson lines, respectively.
Each line carries a left (right) index.
Full circle: Lead-dot hopping vertex $V/\sqrt{N}$, Open circle:  dot-dot hopping vertex $\frac{t_C}{N}$.
The leading order of this selfenergy is $\mathcal{O}(1)+\mathcal{O}(1/N)$}
\label{fig5}
\end{figure}
The NCA fails in describing the Fermi liquid regime at temperatures much lower than $T_K$ 
(again, the NCA gives reliable results down to a fraction of $T_K$) due to the neglect of vertex corrections
in the two-particle correlation functions of pseudo-fermions and slave bosons (see section IVB). 
\par
Typically, the NCA consists in solving a set of self-consistent
equations coupling fermion and boson propagators. 
To lowest order in both vertices we obtain the Baym functional of Fig. 3. 
This functional consists of two terms $\Phi=\Phi_1+\Phi_2$. 
To lowest order in the lead-dot vertex we obtain the functional $\Phi_1$ (Fig. 3.a) 
which is of leading order $\mathcal{O}(1)$ (the order $\mathcal{O}(\frac{1}{N})$ for the vertex $\frac{V}{\sqrt N}$ 
is not skeleton).
The functional $\Phi_2$ (Fig. 3.b) is constructed from the dot-dot vertex and is of 
leading order $\mathcal{O}(\frac{1}{N})$. 
In principle, it is posible to construct another generating functional from the dot-dot vertex 
which contains off-diagonal propagators (this functional can be constructed from $\Phi_2$ by replacing all the 
diagonal fermion and boson propagators by off-diagonal ones).
This functional, however,
does not contribute to leading order with terms $\mathcal{O}(\frac{1}{N})$ in the interdot vertex 
(in other words, there are no off-diagonal selfenergies to second order in
the interdot vertex).                
The reason being that the {\it bare} 
(in the dot-dot coupling sense) 
off-diagonal propagators are zero.
It is, thus, consistent to neglect off-diagonal selfenergies within 
the NCA approximation.
This way, our NCA guarantees that all diagrams of leading order $\mathcal{O}(\frac{1}{N})$ are included within 
a more general subset of diagrams which includes terms to all orders in both vertices.
The NCA solution obtained from $\Phi=\Phi_1+\Phi_2$ is expressed diagrammatically in Figs.\ref{fig4} and 
\ref{fig5}.
These self-energies are obtained by functional derivation
of the Baym functional (Fig. 3) 
$\Sigma=\frac{\partial\Phi}{\partial G}$ and $\Pi=\frac{\partial\Phi}{\partial B}$.
This guarantees that our approximation is conserving.
The obtained selfenergies are of leading order $\mathcal{O}(1)+\mathcal{O}(1/N)$ (bosons) and
$\mathcal{O}(1/N)+\mathcal{O}(1/N^2)$ (fermions).
Applying the real time analytical continuations of Ref. \onlinecite{Lang} to the complex-contour-time-ordered 
fermion self-energy in 
Fig. \ref{fig4} 
we obtain the lesser, greater and retarded components: 
\begin{widetext}
\begin{eqnarray}
\Sigma_{L(R) , \sigma}^{<}(t-t')&=&\frac{1}{N} K_{L(R),\sigma}^{<}(t-t') B_{L(R)}^{<}(t-t')
+(\frac{t_C}{N})^2 G_{R(L),\sigma}^{<}(t-t')B_{R(L)}^{>}(t'-t)B_{L(R)}^{<}(t-t')\nonumber\\
\Sigma_{L(R) , \sigma}^{>}(t-t')
&=&\frac{1}{N} K_{L(R),\sigma}^{>}(t-t') B_{L(R)}^{>}(t-t')
+(\frac{t_C}{N})^2 G_{R(L),\sigma}^{>}(t-t')B_{R(L)}^{<}(t'-t)B_{L(R)}^{>}(t-t')\nonumber\\
\Sigma_{L(R) , \sigma}^{r}(t-t')
&=&\frac{1}{N} \{K_{L(R),\sigma}^{>}(t-t') B_{L(R)}^{r}(t-t')+
K_{L(R),\sigma}^{r}(t-t') B_{L(R)}^{<}(t-t')\}\nonumber\\
&+&(\frac{t_C}{N})^2 \{G_{R(L),\sigma}^{>}(t-t')B_{R(L)}^{<}(t'-t)B_{L(R)}^{r}(t-t')\nonumber\\
&+&G_{R(L),\sigma}^{>}(t-t')B_{R(L)}^{a}(t'-t)B_{L(R)}^{<}(t-t')\nonumber\\
&+&G_{R(L),\sigma}^{r}(t-t')B_{R(L)}^{>}(t'-t)B_{L(R)}^{<}(t-t')\}.
\end{eqnarray}

Where $K_{L(R),\sigma}(t-t')$ is the fermion propagator in the left (right) lead. 
The corresponding expressions for slave-boson self-energies (Fig. 5) are:
\begin{eqnarray}
\Pi_{L(R)}^{<}(t-t')&=&\frac{1}{N}\sum _{\sigma} K_{L(R),\sigma}^{>}(t'-t) G_{L(R) ,\sigma}^{<}(t-t')
+(\frac{t_C}{N})^2  \sum_{\sigma}B_{R(L)}^{<}(t-t')G_{R(L) ,\sigma}^{>}(t'-t)G_{L(R) ,\sigma}^{<}(t-t')\nonumber\\
\Pi_{L(R)}^{>}(t-t')&=&\frac{1}{N}\sum _{\sigma} K_{L(R),\sigma}^{<}(t'-t) G_{L(R) ,\sigma}^{>}(t-t')
+(\frac{t_C}{N})^2  \sum_{\sigma}B_{R(L)}^{>}(t-t')G_{R(L) ,\sigma}^{<}(t'-t)G_{L(R) ,\sigma}^{>}(t-t')\nonumber\\
\Pi_{L(R)}^{r}(t-t')&=&\frac{1}{N}\sum _{\sigma} \{K_{L(R),\sigma}^{<}(t'-t) G_{L(R) ,\sigma}^{r}(t-t')+
K_{L(R),\sigma}^{a}(t'-t) G_{L(R) ,\sigma}^{<}(t-t')\}\nonumber\\
&+&(\frac{t_C}{N})^2 \sum_{\sigma} \{B_{R(L)}^{>}(t-t')
G_{R(L) ,\sigma}^{<}(t'-t)G_{L(R) ,\sigma}^{r}(t-t')\nonumber\\
&-&B_{R(L)}^{r}(t-t')G_{R(L) ,\sigma}^{<}(t'-t)G_{L(R) ,\sigma}^{<}(t-t')\nonumber\\
&+&B_{R(L)}^{<}(t-t')G_{R(L) ,\sigma}^{a}(t'-t)G_{L(R) ,\sigma}^{<}(t-t')\}.
\end{eqnarray}
Eqs. (10) and (11) are the unprojected full NCA self-energies coming from the generating functional.
The projection of these quantities
onto the physical subspace
$\hat{Q}_{\alpha\in \{ L,R \}}=1$ is discussed in Appendix A. 
\end{widetext} 
\subsection{Physical correlation functions}
The physical lesser and greater correlation functions ($\alpha\in \{ L,R \}$) are:
\begin{eqnarray} 
A^{<}_{\alpha\sigma}(t-t')&\equiv&\langle d^\dagger_{\alpha\sigma}(t')
d_{\alpha\sigma}(t)\rangle\nonumber\\
A^{>}_{\alpha\sigma}(t-t')&\equiv&\langle d_{\alpha\sigma}(t)
d^\dagger_{\alpha\sigma}(t')\rangle.
\end{eqnarray}
In terms of slave operators they become the two-particle correlation functions:
\begin{eqnarray}
A^{<}_{\alpha\sigma}(t-t')&\equiv&\langle f^\dagger_{\alpha\sigma}(t')b_\alpha(t')
b^\dagger_\alpha(t) f_{\alpha\sigma}(t)\rangle\nonumber\\
A^{>}_{\alpha\sigma}(t-t')&\equiv&
\langle b^\dagger_\alpha(t) f_{\alpha\sigma}(t)f^\dagger_{\alpha\sigma}(t')b_\alpha(t')\rangle.
\end{eqnarray}
The evaluation of these two-particle correlation functions would require in principle a further 
diagrammatic expansion. 
\begin{figure}
\includegraphics [width=0.25\textwidth,clip]{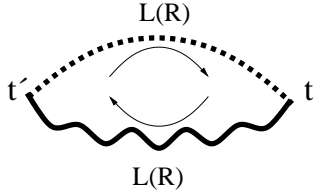}
\caption{Diagrammatic representation of the physical two-particle correlation function within the NCA approximation.
The neglected vertex corrections are $\mathcal{O}(\frac{1}{N^2})$.\cite{slave3}}
\label{fig6}
\end{figure}    
Within the NCA approximation, however, one neglects vertex corrections \cite{vertex}
and keeps only the lowest order term in the expansion of the two-particle correlation function (Fig. 6):
\begin{eqnarray}
A^{<}_{\alpha\sigma}(t-t')&=&\langle f^\dagger_{\alpha\sigma}(t')f_{\alpha\sigma}(t)\rangle
\langle b_\alpha(t')b^\dagger_\alpha(t)\rangle\nonumber\\
&=&G^{<}_{\alpha\sigma}(t-t')B_{\alpha}^{>}(t'-t)\nonumber\\
A^{>}_{\alpha\sigma}(t-t')&=&
\langle b^\dagger_\alpha(t) b_\alpha(t')\rangle 
\langle f_{\alpha\sigma}(t)f^\dagger_{\alpha\sigma}(t')\rangle\nonumber\\
&=&B_{\alpha}^{<}(t'-t)
G^{>}_{\alpha\sigma}(t-t').
\end{eqnarray}
\begin{widetext} 
Using the identities
\begin{eqnarray}
G_{L(R),\sigma}^{>}(t-t')&=&i[G_{L(R),\sigma}^{r}(t-t')-G_{L(R),\sigma}^{a}(t-t')]
-G_{L(R),\sigma}^{<}(t-t')\nonumber\\
B_{L(R),\sigma}^{>}(t-t')&=&i[B_{L(R),\sigma}^{r}(t-t')-B_{L(R),\sigma}^{a}(t-t')]
+B_{L(R),\sigma}^{<}(t-t')
\end{eqnarray}
Eq. (14) can be rewritten as:
\begin{eqnarray}
A^{<}_{\alpha\sigma}(t-t')&=&
G^{<}_{\alpha\sigma}(t-t')\Big\{i[B_{\alpha,\sigma}^{r}(t'-t)-B_{\alpha,\sigma}^{a}(t'-t)]
+B_{\alpha,\sigma}^{<}(t'-t)\Big\}\nonumber\\
A^{>}_{\alpha\sigma}(t-t')&=&
B_{\alpha}^{<}(t'-t)\Big\{i[G_{\alpha,\sigma}^{r}(t-t')-G_{\alpha,\sigma}^{a}(t-t')]
-G_{\alpha,\sigma}^{<}(t-t')\Big\}.
\end{eqnarray}
Now, according to the projection procedure explained in appendix A (see also Ref. \onlinecite{NCAneq1}) the terms 
$G^{<}_{\alpha\sigma}(t-t')B_{\alpha,\sigma}^{<}(t'-t)$
and $B_{\alpha}^{<}(t'-t)G_{\alpha,\sigma}^{<}(t-t')$ have to be projected out (they are of order
$\mathcal{O}(e^{-2i\beta\lambda_\alpha})$). This is accomplished by making:
\begin{eqnarray} 
G_{L(R),\sigma}^{>}(t-t')&=&i[G_{L(R),\sigma}^{r}(t-t')-G_{L(R),\sigma}^{a}(t-t')]
\nonumber\\
B_{L(R),\sigma}^{>}(t-t')&=&i[B_{L(R),\sigma}^{r}(t-t')-B_{L(R),\sigma}^{a}(t-t')]
\end{eqnarray} 
which gives the following physical correlation functions:
\begin{eqnarray}
A^{<}_{\alpha\sigma}(t-t')&=&
iG^{<}_{\alpha\sigma}(t-t')[B_{\alpha,\sigma}^{r}(t'-t)-B_{\alpha,\sigma}^{a}(t'-t)]\nonumber\\
A^{>}_{\alpha\sigma}(t-t')&=&
iB_{\alpha}^{<}(t'-t)[G_{\alpha,\sigma}^{r}(t-t')-G_{\alpha,\sigma}^{a}(t-t')]
\nonumber\\
A^{r(a)}_{\alpha\sigma}(t-t')&=&
G_{\alpha,\sigma}^{r(a)}(t-t')B_{\alpha}^{<}(t'-t)-G^{<}_{\alpha\sigma}(t-t')
B_{\alpha,\sigma}^{a(r)}(t'-t),
\end{eqnarray}
These Green's functions have to be calculated with the corresponding projected selfenergies, 
as discussed in the next section.
\subsection{Physical selfenergies} 
The final set of projected selfenergies is (see Appendix A for details):
\begin{eqnarray}
\Sigma_{L(R) , \sigma}^{r}(\tau)&=&\Bigg\{\frac{1}{N} \tilde{K}_{L(R),\sigma}^{>}(\tau)
+i(\frac{t_C}{N})^2 [\tilde{G}_{R(L),\sigma}^{r}(\tau)-\tilde{G}_{R(L),\sigma}^{a}(\tau)]
\frac{\tilde{B}_{R(L)}^{<}(-\tau)}{Z_{R(L)}}\Bigg\}\times \tilde{B}_{L(R)}^{r}(\tau)\nonumber\\
\Pi_{L(R)}^{r}(\tau)&=&\Bigg\{\frac{1}{N}\sum _{\sigma} \tilde{K}_{L(R),\sigma}^{<}(-\tau)
+i(\frac{t_C}{N})^2 \sum_{\sigma} [\tilde{B}_{R(L)}^{r}(\tau)-\tilde{B}_{R(L)}^{a}(\tau)]
\frac{\tilde{G}_{R(L) ,\sigma}^{<}(-\tau)}{Z_{R(L)}}\Bigg\}\times
\tilde{G}_{L(R) ,\sigma}^{r}(\tau).
\end{eqnarray}  
\begin{eqnarray}
\Sigma_{L(R) , \sigma}^{<}(\tau)
&=&\Bigg\{\frac{1}{N} \tilde{K}_{L(R),\sigma}^{<}(\tau)
+i(\frac{t_C}{N})^2 \frac{\tilde{G}_{R(L),\sigma}^{<}(\tau)}{Z_{R(L)}}
[\tilde{B}_{R(L)}^{r}(-\tau)-\tilde{B}_{R(L)}^{a}(-\tau)]\Bigg\}\times \tilde{B}_{L(R)}^{<}(\tau)\nonumber\\
\Pi_{L(R)}^{<}(\tau)
&=&\Bigg\{\frac{1}{N}\sum _{\sigma}
\tilde{K}_{L(R),\sigma}^{>}(-\tau)
+i(\frac{t_C}{N})^2 \sum _{\sigma} \frac{\tilde{B}_{R(L)}^{<}(\tau)}{Z_{R(L)}}
[\tilde{G}_{R(L) ,\sigma}^{r}(-\tau)-\tilde{G}_{R(L) ,\sigma}^{a}(-\tau)]\Bigg\}\times
\tilde{G}_{L(R) ,\sigma}^{<}(\tau).
\end{eqnarray}   
\end{widetext} 
Where we have introduced the notation $\tau=t-t'$.
The conduction electron propagators $\tilde{K}$ are defined in terms of the Fourier
transforms of the {\it bare} conduction electron
propagators (namely, without dot-lead coupling) as:
$\tilde{K}_{\alpha,\sigma}^{\gtrless}(\epsilon)=2\pi\sum_{k_\alpha}V^2\delta(\epsilon-\epsilon_{k_\alpha})
f_\alpha^{\gtrless}(\epsilon)$,
where $f_\alpha^{<}(\epsilon)=\frac{1}{e^{\beta(\epsilon-\mu_\alpha)}+1}$ is the Fermi function
and $f_\alpha^{>}(\epsilon)=1-f_\alpha^{<}(\epsilon)$ (see Ref. \onlinecite{NCAneq1}).
The Green's functions $\tilde{G}, \tilde{B}$ {\it do not} include inter-dot hopping.
Finally, the factors $Z_L$ and $Z_R$
can be identified with the left and right charges in the absence of inter-dot hopping.
They can be obtained from the left and right charges
of two independent single impurity problems (at different chemical potentials $\mu_L$ and $\mu_R$ respectively) . 
It is important to emphasize two aspects of the projection: 
{\it i)} the simplification of the propagators 
($K\rightarrow \tilde{K}$, $G\rightarrow \tilde{G}$ and $B\rightarrow \tilde{B}$) is {\it required} by the 
projection procedure (see Appendix A) 
and is not an additional approximation; 
{\it ii)} this should not be construed to imply that there is no
inter-dot correction in the slave-particle Green's functions that enter into the
physical correlation functions of Eq. (18).

Eqs. (19-20) constitute the main result of this section. The projected self-energies inserted in the appropriate 
Dyson 
equation, give an overall result in Eq. (18) for the physical correlation functions which has the correct order.  
Of course, in the absence of inter-dot coupling we recover from Eqs. (19-20) two independent sets of NCA equations 
for
the left and right single impurity problems. These equations are in agreement with the ones previously 
obtained in
Refs. \onlinecite{NCAneq2,NCAneq1,NCAneq5}.

The equations for the self-energies, together with the Dyson equations 
for the retarded and lesser propagators and the normalization conditions,
close the set of equations to be solved. We numerically iterate them to convergence. 
\section{Results}
\subsection{Density of states}
\begin{figure}
\includegraphics [width=0.45\textwidth,clip]{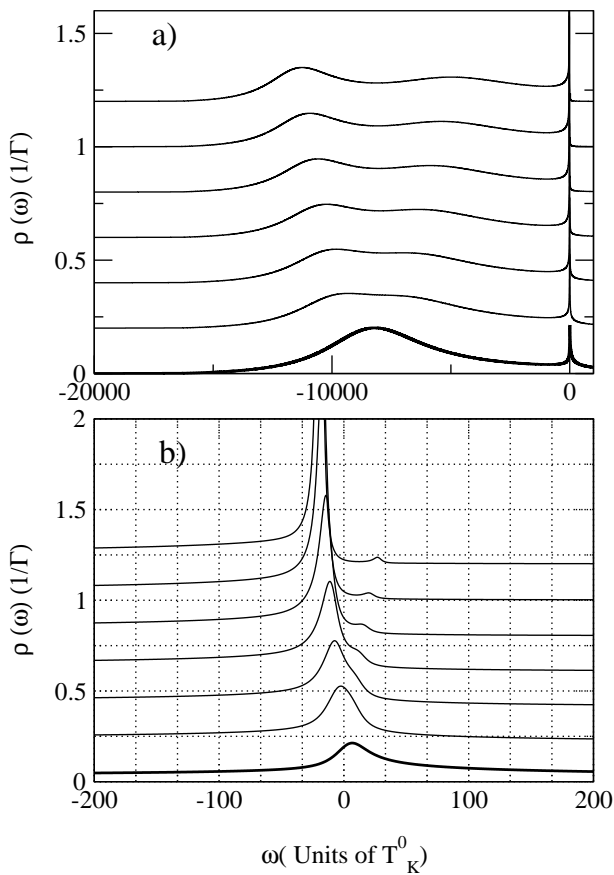}
\caption{Equilibrium density of states (DOS) for different values of the inter-dot hopping 
$t_C=0.0, 1.0, 1.2, 1.4, 1.6, 1.8,$ and $2.0$.
The curves are shifted vertically for clarity. a) Full DOS.
The splitting in the DOS corresponds to the formation
of bonding and anti-bonding combinations of the single particle levels
due to inter-dot tunneling.
b) Blow up of the low frequency region around the Kondo peak. As the inter-dot coupling increases,
the Kondo peak also splits.
Importantly, this splitting,
which is a manifestation of
quantum coherence between the two many body Kondo states on each dot, is much smaller than
the splitting of the broad peak, see main text.}
\label{fig7}
\end{figure}          
Here we present results for the left and right dot densities of states (DOS),
both for equilibrium and finite voltage ($\mu_L=V/2$, $\mu_R=-V/2$) situations. 
\begin{figure}
\includegraphics [width=0.45\textwidth,clip]{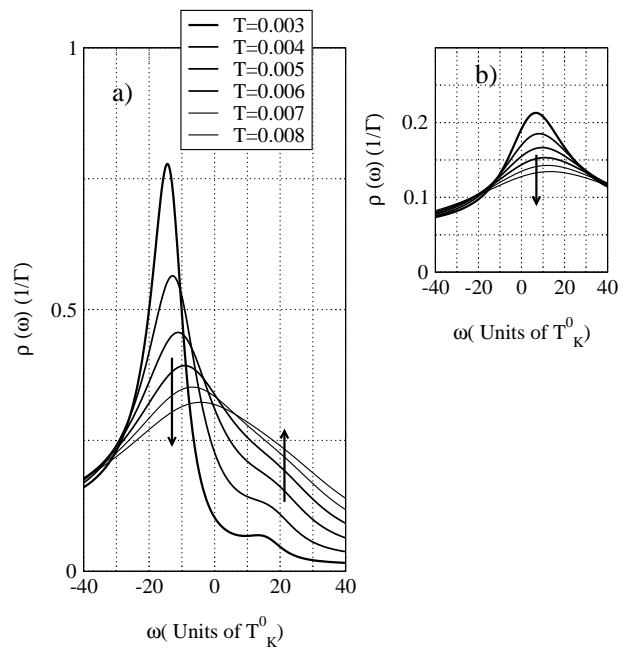}
\caption{Temperature dependence of the equilibrium density of states (DOS) around the Fermi level. The arrows 
indicate
the direction of increasing temperatures. $T_K^0=0.00037$.
Main figure: DOS of the coupled dot problem for $t_C=1.6$.
The structure originating from the inter-dot coupling is still visible at temperatures $T\gtrsim 10 T^0_K$.
The inset shows the reduction of the Kondo peak for the single impurity problem as the temperature increases.
At the highest studied temperature $T\sim 20T^0_K$ the Kondo peak is almost suppressed as
compared with the coupled dot system.}
\label{fig8}
\end{figure}
\begin{figure}
\includegraphics [height=0.5\textwidth,width=0.5\textwidth,clip]{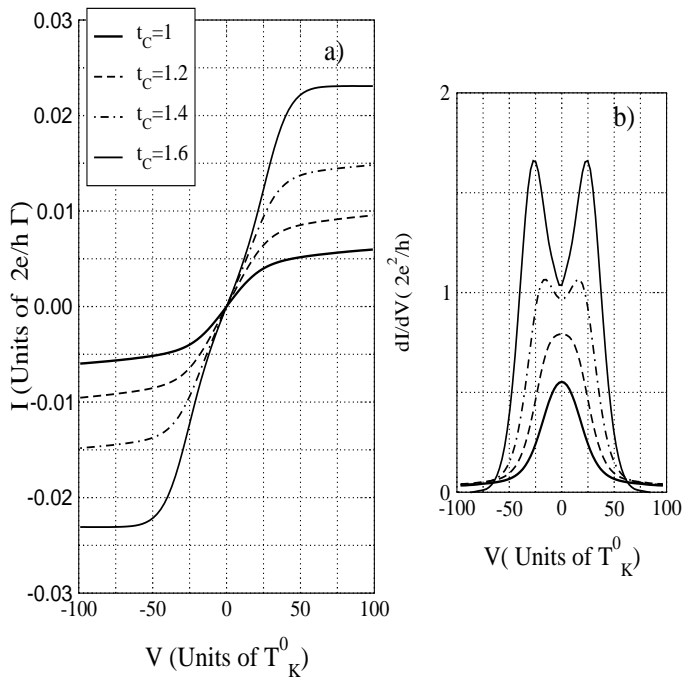}
\caption{Non-linear transport properties of the DQD system for different inter-dot couplings
for $T=0.003\approx 8T_K^0$.
a) Current-voltage characteristics. b) Differential conductance at finite voltage.
The zero-bias anomaly first broadens and then splits with increasing inter-dot hopping.
The splitting of the zero bias
anomaly reflects quantum coherence between the two many-body Kondo states on each QD.}
\label{fig9}
\end{figure}      
We use the following parameters in the calculations (unless otherwise stated):
$\epsilon_0=-2.5$, T=0.003, 
all energies are given in units of
$\Gamma(\epsilon)=2\Delta(\epsilon)=2\pi V_0^2\sum_k \delta(\epsilon-\epsilon_k)=
\pi V^2\sum_k \delta(\epsilon-\epsilon_k)$.
Each lead is described by a parabolic density of states centered at the chemical potential
and with a bandwidth $W=2D=12$.
The Kondo temperature corresponding to these parameters is
$T^0_K\sim 3.7\times 10^{-4}$ (here, the superscript "0" means without inter-dot coupling, 
namely the Kondo temperature of the single impurity problem corresponding to these parameters),
as calculated from the
Bethe ansatz analytical solution ($N=2$) \cite{Hew,NCA2}:
\begin{eqnarray}
T^0_K=\Gamma(1+1/2)D_r(2\Delta/\pi D_r)^{1/2}e^{-\pi|\epsilon_0|/2\Delta}.
\end{eqnarray}
$\Gamma(x)$ is the gamma function and the re-scaling $D_r=e^{-1/2}D$ accounts for
the assumed parabolic DOS in the leads instead of the rectangular one used in the Bethe
ansatz solution \cite{NCAneq1}.                                                    
Note, finally, that in order to compare with the slave-boson mean field (SBMF) results
$T_K^{SBMF}\sim D_r e^{-\pi|\epsilon_0|/2\Delta}\sim 4 T^0_K$.\\
It is known from slave-boson MFA \cite{Aon2,Geor,Agua1} and from numerical renormalization group \cite{Num}
calculations that the effective
Kondo temperature $T_K^{DD}$ of the double dot system grows exponentially with the inter-dot hopping.
Unfortunately, it is not possible to extract an analytical expression for the Kondo temperature from our set of 
coupled NCA equations.
We choose then relatively high temperatures $T>T^0_K$ in all our calculations
in order to prevent the expected low-temperature pathologies
should the effective Kondo temperature $T_K^{DD}$ increase with the inter-dot hopping.\\            
In Fig. 7, the QD density of states (DOS) at equilibrium
(here, of course, the left and right dot are equivalent) is plotted for
increasing values of the inter-dot tunneling.
The full DOS (Fig. 7a) shows the splitting of the main peak 
(energy scale for charge fluctuations) originating from the inter-dot coupling which 
generates quantum coherence between the dots.
The splitting in the DOS corresponds to the formation
of bonding and anti-bonding combinations of the single particle levels, i.e.
$\epsilon_{\pm}=\epsilon_{0}\mp t_C$ due to inter-dot tunneling.
Fig. 7b, shows a blow up of the low frequency part of the DOS around the Fermi level. 
As we increase the inter-dot coupling, the Kondo peak also splits into bonding and anti-bonding combinations.
Importantly, the energy scale for this splitting of the Kondo peak, 
which is a manifestation of 
quantum coherence between the two many body Kondo states on each dot, is much smaller than
the one corresponding to the splitting of the broad peak 
(which is a manifestation of coherence between single particle states). We have, then, 
$\Delta\tilde{\epsilon}=2\tilde{t}_C <<\Delta\epsilon=2t_C$, where $\Delta\tilde{\epsilon}$ and 
$\Delta\epsilon=2t_C$ are the splitting of the Kondo peak and the single particle splitting, respectively.
This reduction of the splitting, namely $\tilde{t}_C<<t_C$, is caused by the strong Coulomb repulsion on each dot.
Typical values of this splitting are in the range $\sim 10T^0_K-40T^0_K$ (note that the single particle splittings
are in the range $\sim 10^3T^0_K-10^4T^0_K$). 
These obtained values for the reduced splitting of the Kondo resonance 
are in good semiquantitative agreement
with the mean-field slave boson calculation.\cite{Aon,Aon2,Geor,Agua1}

The behavior at different temperatures is studied in Fig. 8 where we plot
the DOS of the coupled dot problem (Fig. 8a) for $t_C=1.6$ at different temperatures.
The splitting originating from the inter-dot coupling is still visible at temperatures $T\gtrsim 10 T^0_K$.
For comparison, we show in Fig. 8b the reduction of the Kondo peak for the single impurity problem as the 
temperature increases.
At the highest studied temperature $T\sim 20T^0_K$ the Kondo peak for the single impurity system is almost 
suppressed as 
compared with the coupled dot system.
This is in good qualitative agreement with the previous statement that $T_K^{DD}>T^0_K$ \cite{Aon2,Geor,Agua1,Num}.
It is worth noting that the splitting of the Kondo resonance is robust at temperatures 
higher than $T^0_K$; experimentally this is of the most relevance:
according to this result, the experimental conditions for studying Kondo physics in coupled QD's are 
less demanding than in single QD's
(temperatures much lower than $T^0_K$ are needed in order to observe Kondo-related
features in the transport properties of single QD's \cite{Gold,Leo,Stutt,Blick,Wilfred}).  
\subsection{Non-linear transport properties}
We have proven in the previous section that the inter-dot coupling generates quantum coherence between the dots. This quantum coherence
is reflected in the DOS of each QD as a splitting, {\it both in the charge fluctuation and spin fluctuation
parts of the spectrum}. 
We are interested in Kondo physics and the obvious question we want to answer is thus: 
Can we observe the splitting of the Kondo peak, induced by the inter-dot coupling,
in a differential conductance measurement? 
The answer to the previous question is non-trivial because we are dealing with
the non-equilibrium physics of strongly correlated electrons and hence
the spectral functions are expected to strongly depend
on the applied bias voltage (shift and broadening of the peaks).
In other words, the differential conductance curve does not just mimic the
zero-voltage DOS (as it does for non-interacting electrons). 
From the experimental point of view this is a timely and crucial question: 
the observation of such a splitting would prove the remarkable phenomenon of
quantum-coherence between {\it the two Kondo many-body states on each dot}. 
Experiments by Oosterkamp {\it et al} \cite{Oos} and Blick {\it et al} \cite{Blick2} 
have proven quantum coherence between single particle states in coupled QD's.
Also, some signatures of coherence between Kondo states in a double quantum dot system have been reported 
recently by Jeong {\it et al} in Ref. \onlinecite{Chang}.
The first step in order to answer our question is to calculate the current through the double dot system. 
We follow
the standard non-equilibrium approach to transport through a region of interacting electrons 
\cite{Hers1,Win2} and relate the current through 
each dot to its retarded and lesser Green's functions:
\begin{eqnarray}
I_{\alpha\in \{ L,R \}}=-\frac{2e}{h}\int d\epsilon 
\Gamma(\epsilon)[2Im A^r_{\alpha}(\epsilon)f_\alpha(\epsilon)
+A^<_{\alpha}(\epsilon)].\nonumber\\ 
\end{eqnarray}
Here, $A^r_{\alpha}(\epsilon)$ and $A^<_{\alpha}(\epsilon)$ are the Fourier transforms
of the retarded and lesser physical Green's functions of Eq. (18).
\begin{figure}
\includegraphics [height=0.4\textwidth,width=0.4\textwidth,clip]{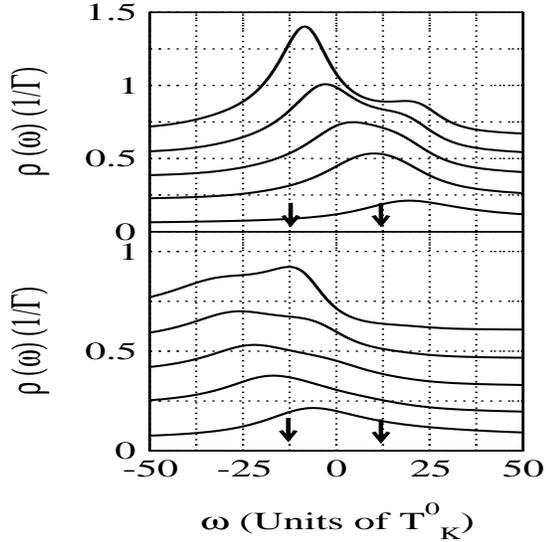}
\caption{Non-equilibrium DOS ($\mu_L=V/2=12.5T^0_K$, $\mu_R=-V/2=-12.5T^0_K$)
for different values of the inter-dot hopping $t_C=0.0, 1.0, 1.2, 1.4,$ and $1.6$. The curves are shifted vertically
for clarity.
Top: left DOS. Bottom: Right DOS. The arrows mark the position of the chemical potentials.}
\label{fig10}
\end{figure}
\begin{figure}
\includegraphics [height=0.4\textwidth,width=0.4\textwidth,clip]{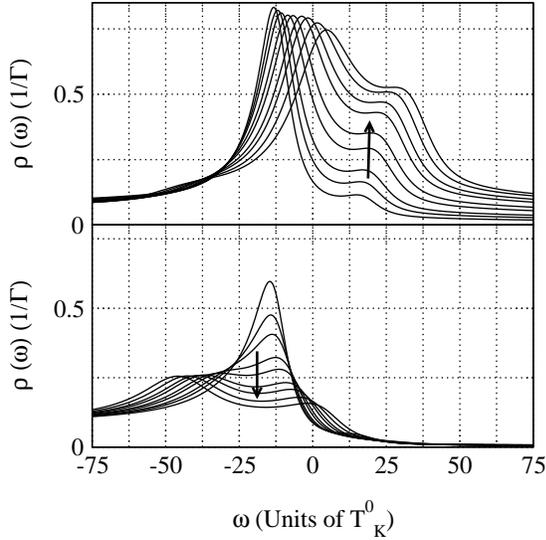}
\caption{Non-equilibrium DOS at $t_C=1.6$
for different voltages in the range $V=10T^0_K$ through $50T^0_K$.
Top: left DOS. Bottom: Right DOS. The arrows mark the directions of increasing voltages.}
\label{fig11}
\end{figure}      
The total current through the system is calculated as $I=\frac{(I_L-I_R)}{2}$.
The differential conductance ${\cal G}=dI/dV$ is calculated by numerical differentiation 
of the current-voltage (I-V) curves.

We study in Fig. 9 the non-linear transport properties of the DQD system. 
We plot in Fig. 9a 
the I-V characteristics for different values of the inter-dot hopping. As the inter-dot hopping increases, 
the low-voltage differential conductance grows 
while at the same time 
starts to deviate from an Ohmic behavior. At large voltages the current saturates, the differential 
conductance nears zero 
and even becomes slightly negative for the largest $t_C$. 
\begin{figure}
\includegraphics [height=0.5\textwidth,width=0.5\textwidth,clip]{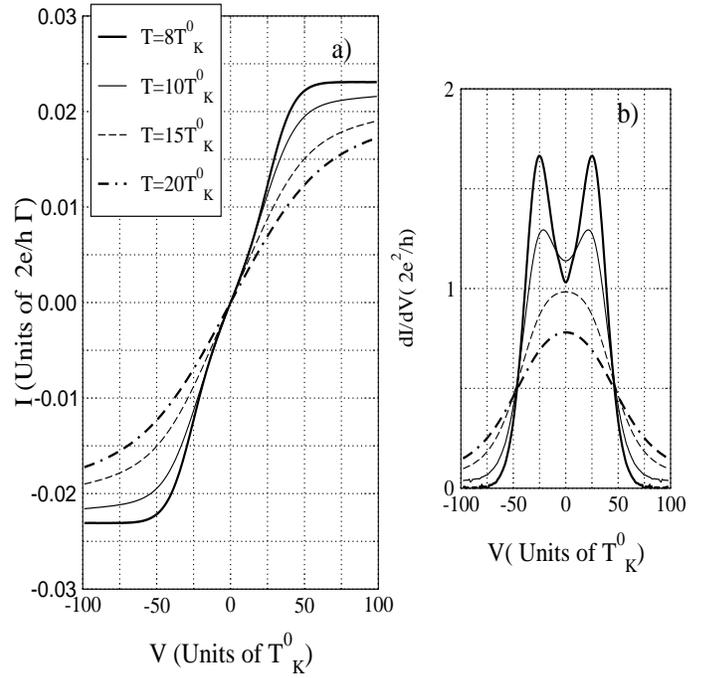}
\caption{Non-linear transport properties of the DQD system ($t_C=1.6$) for different temperatures as
a function of the applied bias voltage.
a) Current-voltage characteristics. b) Differential conductance at finite voltage.}
\label{fig12}
\end{figure}
\begin{figure}
\includegraphics [height=0.5\textwidth,width=0.5\textwidth,clip]{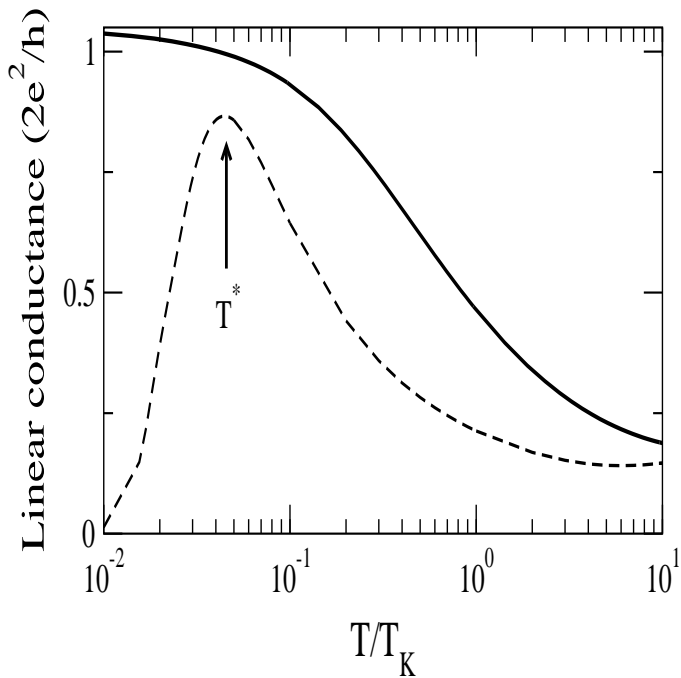}
\caption{Comparison of the temperature dependence of the linear conductance in a single quantum dot (solid line)
and a double quantum dot with $t_C=1.6$ (dashed line).
The linear conductance for the single dot follows the usual logarithmic increase at intermediate temperatures followed by a
saturation near the unitary limit. The linear conductance for the double dot case shows a nonmonotonic temperature dependence,
it increases for decreasing temperatures in the region $T>T^*$ whereas it decreases in the region $T<T^*$.
The temperature scale $T^*$ characterizes quantum coherence between both dots in the
Kondo regime. Note that in order to compare with the single dot case the temperature has been scaled with respect to the
Kondo temperature of the {\it single dot problem}, see main text.}
\label{fig13}
\end{figure}       
These features are better brought out in a plot of the differential
conductance at finite voltage (Fig. 9b). As we increase the inter-dot hopping, the 
zero bias Kondo anomaly 
broadens and splits. We can attribute this broadening to the aforementioned increase of the effective Kondo temperature 
as a function of the inter-dot hopping. For large inter-dot tunneling couplings the zero-bias anomaly splits. 
The splitting of the zero bias anomaly 
is an unambiguous indication of quantum coherence between the Kondo states on each dot.

In Fig. 10 we plot the non-equilibrium DOS ($\mu_L=V/2=12.5T^0_K$, $\mu_R=-V/2=-12.5T^0_K$) 
for the left (top figure) and right (bottom figure) coupled quantum dots.
For the uncoupled situation ($t_C=0$) each DOS has a Kondo peak around each chemical potential as expected.
With increasing inter-dot hopping, the behavior of each DOS becomes quite complex. 
The Kondo peak on each side splits into two peaks while at the same time
the whole spectral weight near the Fermi level shifts to lower frequencies.
Furthermore, these split peaks are asymmetric, they have different heights and spectral weights
(it is important to mention here, however, that the NCA is known to overestimate 
the asymmetry of the peaks because
it does incorrectly treat potential and spin flip scattering on equal footing \cite{Kroha2}).
As the inter-dot hopping increases, the lower (upper) band of the left (right) DOS moves 
to lower (higher) frequencies, while increasing its height, until it matches with the upper (lower) 
band of the right (left) DOS. 
As an example, for $t_C=1.6$ the lower peak on the left DOS and the upper peak on the right DOS
approximately match at $\omega\sim \mu_R$. As a result, there is a peak in the 
differential conductance at $V=25T^0_K$ for 
this inter-dot coupling.\\ 
Also interesting is to study how the DOS evolves as a function of the applied voltage for a fixed inter-dot 
coupling.
This non-trivial behavior of the DOS versus applied voltage is studied in Fig. 11
where we plot the non-equilibrium DOS for $t_C=1.6$ and different
voltages from $V=10T^0_K$ to $V=50T^0_K$ in intervals of $\Delta V=5T^0_K$.
As the voltage increases,
the left (right) DOS moves to higher (lower) frequencies such that the middle point between the
split Kondo peaks
lies approximately at the left (right) chemical potential (this discussion is only qualitative,
note that even for the uncoupled
case the Kondo peaks {\it do not} lie exactly on each chemical potential).

The temperature dependence of the current and differential conductance are plotted in Fig. 12.
Several features in these curves are noteworthy. If we focus first in the differential conductance (Fig. 12.b) 
we see that the splitting of the zero-bias anomaly can be resolved for temperatures $T\alt 10 T^0_K$. For higher
temperatures the splitting can no longer be resolved and, instead, a broad zero-bias anomaly is obtained.
Also important to mention is the non-monotonic behavior of the linear conductance 
${\cal G}=dI/dV|_{V=0}$ with temperature. 
Starting from high temperatures, the linear conductance first {\it increases}
for decreasing temperatures, indicating the
appearance of Kondo physics. This behavior saturates at the temperature for which the splitting is 
resolved (here
$T\sim 10 T^0_K$) and then the linear conductance {\it decreases} for decreasing temperatures.
This behavior can
be easily explained by noting that the linear conductance at finite temperatures is a convolution of the
DOS around the Fermi level with the derivative of the Fermi function
(whose full-width at half-maximum is 3.5T). When the width of the derivative of the Fermi function is smaller
than the splitting of the Kondo peak this convolution is very small, due to the small spectral weight
around the Fermi level when the Kondo peak splits, explaining why the linear conductance decreases 
when lowering the 
temperature. This non-monotonic temperature behavior is an indirect proof of the formation of the
splitting (in single dots in the Kondo regime the linear conductance monotonicaly increases,
until it saturates in the Fermi liquid regime, for decreasing temperatures).
We show this behavior in Fig. 13 where we compare the temperature dependence of the linear conductance 
of a single quantum dot (solid line)
with the temperature dependence of the linear conductance of a double quantum dot with $t_C=1.6$ (dashed line).
The linear conductance for the single dot follows the usual logarithmic increase at intermediate temperatures followed by a
saturation near the unitary limit. The linear conductance for the double dot case shows a nonmonotonic temperature dependence,
it increases for decreasing temperatures in the region $T>T^*$ whereas it decreases in the region $T<T^*$.
The temperature scale $T^*$ (which is the temperature for which the splitting is
resolved in Fig. 12. b) characterizes quantum coherence between both dots in the
Kondo regime. Note that in order to compare with the single dot case the temperature has been scaled with respect to the
Kondo temperature of the {\it single dot problem}, $T_K=D\sqrt{2\Delta/\pi|\epsilon_0|}e^{-\pi|\epsilon_0|/4\Delta}$.
Finally, it is important to mention here is that the NCA is known to overestimate the Kondo peak 
amplitude 
(and then the linear conductance) when 
calculated from the density of states. 
Typical overestimates are within the range $10-15\%$ \cite{NCAneq2}. 
Keeping this overestimation in mind (which for temperatures
$T\lesssim 4\times 10^{-2} T_K$ leads to an overshooting of the unitary limit in the single dot case, 
Fig. 13 solid line), 
we purposefully show results at low temperatures
where the temperature dependences of the linear conductance for single and double dot 
cases compare best.
\begin{figure}
\includegraphics [height=0.5\textwidth,width=0.5\textwidth,clip]{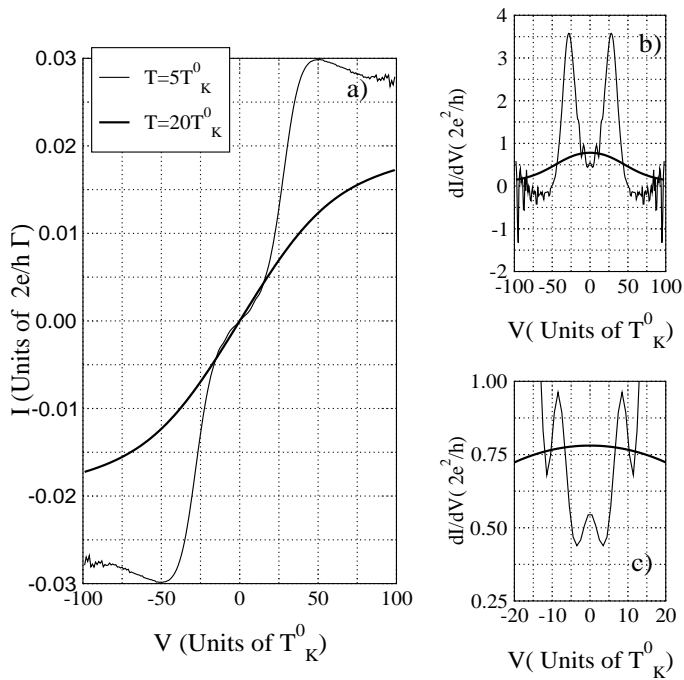}
\caption{Non-linear transport properties of the DQD system ($t_C=1.6$) for two different temperatures as
a function of the applied bias voltage.
a) Current-voltage characteristics. b) Differential conductance at finite voltage. 
At large voltages the system
develops regions of negative differential conductance accompanied by fluctuations in the current. 
We speculate that these fluctuations could originate from a dynamical instability in this region 
of voltages.
c) Blow-up of the low voltage region in the differential conductance. The extra structure at low voltages 
(small zero-bias anomaly+satellites) is originating from the splitting due to the applied voltage.}
\label{fig14}
\end{figure} 

Finally, we comment on the temperature dependence of the differential conductance at large 
voltages (see Fig. 12.a). 
At low temperatures the slope of the I-V characteristics at large voltages approaches zero and eventually becomes
slightly negative, namely the I-V characteristics present negative differential conductance (NDC), at the 
lowest temperatures. The slope of the differential conductance increases gradually 
as one increases the temperature. For the highest temperature studied ($T=20T_K^0$, dashed-dotted line) no traces
of NDC are found even for very large voltages.

We study this NDC behavior in
Fig. 14, where we compare the I-V characteristics (Fig. 14a) and differential conductance (Fig. 14b) 
of the system at high 
($T=20T^0_K$) and low temperatures ($T=5T^0_K$). For the low temperature situation, 
the slope at large voltages
does indeed develop NDC for voltages $V\agt 50 T^0_K$. At $V\simeq 50 T^0_K$ the differential conductance becomes zero and the current 
smoothly starts to decrease as one increases the dc voltage. However, the situation changes drastically at larger dc 
voltages where our numerical results for the current rapidly develop a wiggly pattern.
The appearance of this fluctuating pattern in the numerics is accompanied by a breakdown of current conservation, 
namely $J_L=-J_R$ is no longer fulfilled. 
We do not have a conclusive answer for the appearance of this pattern in the current, but, nonetheless, the fact
that it appears in the NDC region together with a breakdown of the condition $J_L+J_R=0$,
allows us to speculate that it may be reflecting
the formation of a dynamical instability where a {\it time dependent} current spontaneously 
develops in response to the
{\it static} applied voltage
(with a non-zero displacement current $J_{disp}(t)=-(J_L+J_R)/2=-\frac{e}{2}\frac{dQ_{DD}}{dt}$, 
where $Q_{DD}$ is the charge accumulated in the 
double dot).\cite{broken} 
Although the analysis of time-dependent phenomena is beyond the purpose
of this work (our numerical scheme is only valid for time-translational invariant situations), 
we just mention that
this kind of dynamical instability, rather typical in non-linear systems presenting 
NDC \cite {NDC},
have been recently reported in single QD's in the Kondo regime.\cite{time1,time2,time3} 
It is, thus, reasonable to expect similar dynamical instabilities in DQD's, which motivates 
our speculation.

We finish this part with two remarks. 
The first  is that this NDC has been previously reported in the context of SBMFT.\cite{Agua1}
Importantly, the NDC features obtained here are smooth 
(the dI/dV evolves from zero to negative values in a smooth manner) and gradually dissapear as the 
temperature increases 
(as already anticipated by us in Ref. \onlinecite{Agua1})
in contrast with the ones obtained within the SBMFT (sharp transitions
between the high and low current regions). 
These sharp transitions can be attributed to the lack of fluctuations (quantum and thermal)
of the boson fields in the SBMFT. \cite{Agua1}

The second remark is that the low-voltage part of the differential conductance curve at the lowest temperature 
(Fig. 14c) does also develop new fine structure (extra peaks). 
The differential conductance develops a small zero-bias anomaly and satellites 
separated from zero at $\Delta V\sim \pm 10T^0_K$. These new structures in the differential conductance 
are in agreement with the ones 
previously reported in two-level
quantum dots \cite{Inos,Poj} and coupled quantum dots in the limit of strong inter-dot repulsion \cite{Poj} 
and can be attributed
to the extra splitting induced by the applied voltage: the voltage splits the peaks in the left and right
spectral functions, a peak in the differential conductance occurs when these split peaks cross each other.
The agreement in only qualitative though.
In Refs. \onlinecite{Inos,Poj} such crossings occur at $\Delta\epsilon=V$ where 
$\Delta\epsilon$ (a fixed quantity)
is the energy 
separation between {\it single particle} levels in the two-level quantum dot \cite{Inos,Poj} or 
the the energy separation between the bonding and anti-bonding levels in the
coupled quantum dot system.\cite{Poj} 
On the contrary, the peaks in the differential conductance of our calculation appear at much lower frequency scales.
As mentioned before, our calculation includes the strong renormalization of the levels due to electronic 
correlations and due to the voltage. The crossings, hence, appear at voltages for which 
$\Delta\tilde{\epsilon} (V)=V$ (namely,
$\tilde{\epsilon}_{+}+V/2=\tilde{\epsilon}_{-}-V/2$),
where $\Delta\tilde{\epsilon} (V)=\tilde{\epsilon}_{-}-\tilde{\epsilon}_{+}=2\tilde{t}_C(V)$ is
the voltage-dependent energy separation between the anti-bonding and bonding combinations of the Kondo peak
(which, again, is much smaller than the single-particle splitting $2t_C$).

\begin{figure}
\includegraphics [height=0.4\textwidth,width=0.4\textwidth,clip]{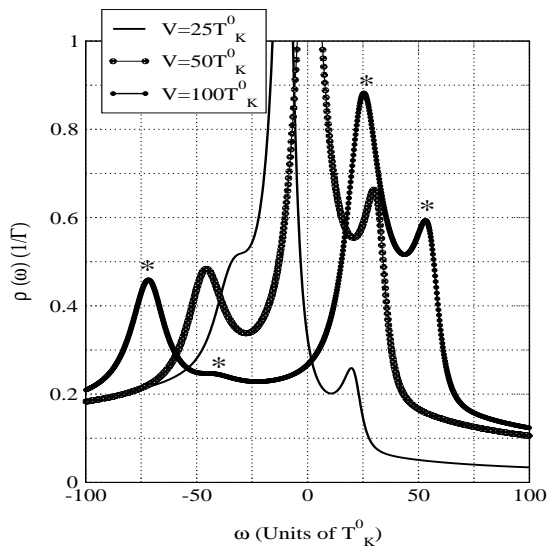}
\caption{Non-equilibrium full DOS at low temperature ($T=5T^0_K$) and $t_C=1.6$
for different voltages $V=25T^0_K,50T^0_K,100T^0_K$. The applied voltage induces extra 
splittings in the
bonding and anti-bonding combination of the Kondo peak. As a result, four peaks can be 
clearly resolved in 
the full DOS at high voltages 
(these peaks are marked with asterisks for the highest voltage in the figure).}
\label{fig15}
\end{figure} 

Fig. 15, where we plot the full spectral function 
at a finite voltage, illustrates this phenomenon.
Each peak splits by $\pm V/2$. As a result the full DOS
develops four peaks, the combinations $\epsilon_+\pm V/2$ and $\epsilon_-\pm V/2$, 
that can be clearly resolved at high enough voltages.
These split peaks are marked with asterisks for the highest voltage in the figure,
the distance between consecutive peaks is twice the renormalized inter-dot hopping, the distance between
alternate peaks is the voltage. We mention, in passing, that the observation of this fine structure
in the differential conductance would constitute a direct proof of the voltage-induced splitting of the 
Kondo resonance. Here, the splitting associated with the inter-dot hopping serves as a 
testing tool, similarly to that provided by an external magnetic field in single quantum dots \cite{Win1} 
(the quantity $\Delta\epsilon$ of our previous discussion being now the zeeman splitting in a single 
quantum dot with an external magnetic field) to check the voltage-induced splitting. 
Different proposals for measuring this voltage-induced splitting are the subject of current 
active research. \cite{NCAneq4,Sun,Silvano}

\begin{figure}
\includegraphics [height=0.4\textwidth,width=0.4\textwidth,clip]{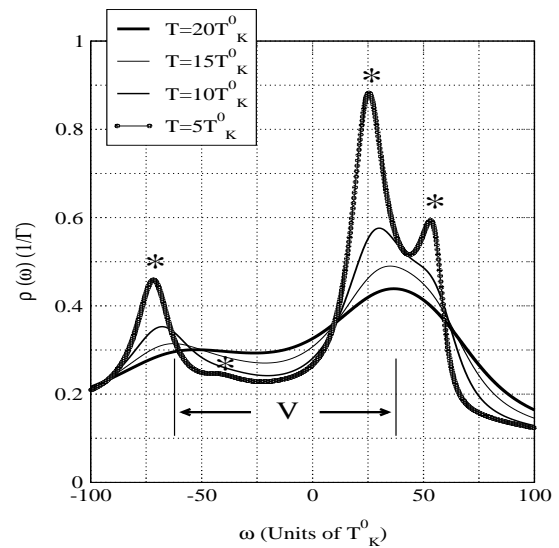}
\caption{Temperature dependence of the non-equilibrium full DOS ($V=100T^0_K$) and $t_C=1.6$.
At high temperatures, the splitting coming from inter-dot coupling can not be resolved and 
the coupled dot system is equivalent to a single dot with a broad Kondo peak
coming from a convolution of the bonding and anti-bonding peaks with thermal broadening. 
This effective single Kondo peak
is split by the voltage as expected (vertical marks). Further lowering of the temperature allows 
the resolution of the inter-dot-induced splitting (asterisks).}
\label{fig16}
\end{figure}  
We support our previous paragraph by studying the temperature dependence of the non-equilibrium full DOS
at $V=100T^0_K$ (Fig. 16). 
At high temperatures, $T\agt 2\tilde{t}_C$, the splitting coming from inter-dot coupling 
can not be resolved and
the coupled dot system is equivalent to a single dot with a broad Kondo peak
(coming from a convolution of the bonding and anti-bonding peaks with thermal broadening). 
The width of this effective Kondo peak is thus larger than $2\tilde{t}_C$.
As expected, a finite voltage, $V>T$, splits this 
effective single Kondo peak into two peaks separated by V (Fig. 16, thick solid line). 
Further lowering of the temperature allows to
resolve the inter-dot-induced splitting resulting in extra peaks superimposed to the ones induced by 
the voltage (Fig. 16, asterisks).\\
We finish by commenting on the observability of the effects described in this section: 
We obtain in our calculations splittings in the differential conductance of the order
of $\simeq 50 T_K^0$. Typical Kondo temperatures in quantum dots are 
of the order of a few $\mu eV$ (for instance, the Kondo temperature is $\simeq 4-250 \mu eV$ in the experiment
of Ref. \onlinecite{Gold}), which gives splittings well within the resolution
limits of state-of-the-art techniques (remember that $1\mu eV\sim 10mK$). 

\section{Conclusions}
We have theoretically studied the transport properties, 
both equilibrium and out-of-equilibrium properties, of a coupled quantum
dot system in the Kondo regime. 
We have modeled the double quantum
dot system by means of a two-impurity Anderson Hamiltonian with inter-impurity hopping
and infinite on-site interaction on each dot.
The Hamiltonian, formulated in slave-boson language, is solved by means of a
generalization of the non-crossing approximation (NCA) for the present problem:
two quantum dots in the Kondo regime, coupled
to each other by a tunneling barrier and with an applied voltage
across them. We have provided benchmark calculations of the predictions of the non-crossing 
approximation for the linear and nonlinear transport properties of coupled quantum dots in the Kondo regime.
We give a series of predictions that can be observed experimentally
in linear and nonlinear transport measurements through coupled quantum dots in the Kondo regime:\\
{\it i)} The nonlinear differential conductance ${\cal G}=dI/dV$ directly measures the transition (as $t_C$ increases)
from two isolated Kondo impurities to a coherent bonding and anti-bonding
superposition of the many-body Kondo states of each dot.
For increasing inter-dot couplings, the zero-bias anomaly first broadens and then splits.
The later case corresponds to transport which is optimized
for a finite bias voltage matching the splitting between these two bonding
and anti-bonding states.\\  
{\it ii)} The effective Kondo temperature of the coupled system increases with the inter-dot coupling.
This is reflected as broadening of the zero-bias anomaly.\\
{\it iii)} The non-monotonic temperature behavior of the linear conductance ${\cal G}=dI/dV|_{V=0}$ 
is an indirect proof of the formation of the
splitting. Starting from high temperatures, the linear conductance first grows
for decreasing temperatures, indicating the appearance of Kondo physics. 
This behavior saturates at the temperature for which the splitting is resolved. 
Further lowering of the temperature produces a decrease of the linear conductance:
The curve linear conductance vs temperature has a maximum at a temperature scale $T^*$ characterizing quantum 
coherence between the two quantum dots.\\
{\it iv)} The differential conductance at large voltages can become negative (NDC). 
We speculate
that the system can develop dynamical instabilities around this region.\\
{\it v)} At low enough temperatures, it is possible to resolve extra structures in the differential
conductance coming from the splitting induced by the applied bias voltage.\\      
We hope our work will inspire and encourage experimental investigations of Kondo physics in 
coupled quantum dots and related systems.
\begin{acknowledgements}
We are thankful to Piers Coleman, Chris Hooley and Olivier Parcollet for many useful and 
stimulating discussions
during the first stages of this work. This work was supported by the NSF
grant DMR 00-93079, DOE grant DE-FG02-99ER45970 and by the MEC of Spain grant PF 98-07497938.  
Ram\'on Aguado acknowledges support from the MCYT of Spain through the "Ram\'on y Cajal" program for 
young researchers.
\end{acknowledgements}
\appendix
\begin{widetext}
\section{Projection}
Here we discuss the evaluation of operator averages within the restricted subspace
of the Hilbert space with the constraints of Eq. (3).
The formal expression for the expectation value of an operator
in this subspace can be written as:
\begin{eqnarray}
\langle \hat{O} \rangle_{Q_L=1,Q_R=1}=\frac{1}{Z_{Q_L=1,Q_R=1}}Tr\{
e^{-\beta(H_0-\mu_L N_L-\mu_R N_R)}\delta_{Q_L,1}\delta_{Q_R,1}
T_C[S_C(-\infty,\infty)\hat{O}]\},
\end{eqnarray}
where $T_C$ orders operators along a complex contour,
the hopping terms are treated as perturbations
(i.e $H_0$ contains the isolated regions of the problem, leads and dots,
before they are connected),
and the partition function is given by:
\begin{eqnarray}
Z_{Q_L=1,Q_R=1}=Tr\{
e^{-\beta(H_0-\mu_L N_L-\mu_R N_R)}\delta_{Q_L,1}\delta_{Q_R,1}
T_C[S_C(-\infty,\infty)]\}.
\end{eqnarray}
Since the charge operators commute with the Hamiltonian
each constraint can be incorporated by a Kronecker delta function in the statistical
averages of Eqs. (A1-A2).
To relate averages in the restricted ensemble with the ones corresponding
to an unrestricted ensemble we represent each Kr\"onecker delta function
as an integral
over a complex chemical potential \cite{NCAneq2,Read} (see also appendix D in Ref. \onlinecite{NCA2} and chapter 7 of 
Ref. \onlinecite{Hew}):
\begin{eqnarray}
\delta_{Q_L,1}&=&\frac{\beta}{2\pi}\int_{-\frac{\pi}{\beta}}^{\frac{\pi}{\beta}}d\lambda_L
e^{-i\beta\lambda_L(Q_L-1)}\nonumber\\
\delta_{Q_R,1}&=&\frac{\beta}{2\pi}\int_{-\frac{\pi}{\beta}}^{\frac{\pi}{\beta}}d\lambda_R
e^{-i\beta\lambda_R(Q_R-1)}.
\end{eqnarray}
\begin{eqnarray}
\langle \hat{O} \rangle_{Q_L=1,Q_R=1}=\frac{1}{Z_{Q_L=1,Q_R=1}}
(\frac{\beta}{2\pi})^2
\int_{-\frac{\pi}{\beta}}^{\frac{\pi}{\beta}}d\lambda_L
\int_{-\frac{\pi}{\beta}}^{\frac{\pi}{\beta}}d\lambda_R
 e^{i\beta\lambda_L}e^{i\beta\lambda_R} Z_{GC} \langle \hat{O} \rangle_{GC}.
\end{eqnarray}      
This way, we can relate the average in the constrained ensemble with an average in the grand canonical ensemble which can be written as:
\begin{eqnarray}
\langle \hat{O} \rangle_{GC}&=&\frac{1}{Z_{GC}}Tr\{
e^{-\beta(H_0-\mu_L N_L-\mu_R N_R+i\lambda_L Q_L+i\lambda_R Q_R)}
T_C[S_C(-\infty,\infty)\hat{O}]\},\nonumber\\
Z_{GC}&=&Tr\{
e^{-\beta(H_0-\mu_L N_L-\mu_R N_R+i\lambda_L Q_L+i\lambda_R Q_R)}
T_C[S_C(-\infty,\infty)]\}.
\end{eqnarray}
This average inside the integral in Eq. (A4) now obeys a linked cluster
theorem and we can use conventional field theory.
In principle, we can stop here, evaluate the averages in the grand canonical ensemble
and projecting to the physical ensemble by a final integration over the chemical potentials.
Further simplification can be gained, however, by noting that
the grand canonical partition function $Z_{GC}$
can be rewritten as a sum over canonical partition functions:
\begin{eqnarray}
Z_{GC}=\sum_{Q_L=0}^{\infty}\sum_{Q_R=0}^{\infty}Z_C(Q_L,Q_R)e^{-i\beta\lambda_L Q_L}
e^{-i\beta\lambda_R Q_R};
\end{eqnarray}
and by expanding the expressions in the grand canonical ensemble as power series:
\begin{eqnarray}
Z_{GC}&=&Z_C(0,0)+Z_C(1,0)e^{-i\beta\lambda_L}+Z_C(0,1)e^{-i\beta\lambda_R}
+Z_C(1,1)e^{-i\beta\lambda_L}e^{-i\beta\lambda_R}+...\nonumber\\
\langle \hat{O} \rangle_{GC}&=&\langle \hat{O} \rangle^{0,0}+
\langle \hat{O} \rangle^{1,0}e^{-i\beta\lambda_L}+
\langle \hat{O} \rangle^{0,1}e^{-i\beta\lambda_R}+
\langle \hat{O} \rangle^{1,1}e^{-i\beta\lambda_L}e^{-i\beta\lambda_R}+...
\end{eqnarray}
Inserting these power series expansions inside the integral in Eq. (A4), the only terms that
survive are:
\begin{eqnarray}
\langle \hat{O} \rangle_{Q_L=1,Q_R=1}=\frac{1}{Z_C(1,1)}
[Z_C(0,0)\langle \hat{O} \rangle^{1,1}+Z_C(1,0)\langle \hat{O} \rangle^{0,1}
+Z_C(0,1)\langle \hat{O} \rangle^{1,0}],
\end{eqnarray}
\end{widetext}                         
where we have used $\langle \hat{O} \rangle^{0,0}=0$ which is the case for any
physical operator of interest. The operators we are interested in can be classified as operators
acting on the left dot or operators acting on the right dot \cite{projection}, namely:
\begin{eqnarray}
\langle \hat{O}_L \rangle_{Q_L=1,Q_R=1}&=&[\frac{Z_C(0,1)}{Z_C(1,1)}\langle \hat{O}_L \rangle^{1,0}
+\frac{Z_C(0,0)}{Z_C(1,1)}\langle \hat{O}_L \rangle^{1,1}]\nonumber\\
\langle \hat{O}_R \rangle_{Q_L=1,Q_R=1}&=&[\frac{Z_C(1,0)}{Z_C(1,1)}\langle \hat{O}_R \rangle^{0,1}
+\frac{Z_C(0,0)}{Z_C(1,1)}\langle \hat{O}_R \rangle^{1,1}].\nonumber\\
\end{eqnarray}
We can conclude from this analysis that physical operators on the
left and right sides
have to be of order
$\mathcal{O}(e^{-i\beta\lambda_L})+\mathcal{O}(e^{-i\beta\lambda_L}e^{-i\beta\lambda_R})$
and $\mathcal{O}(e^{-i\beta\lambda_R})+\mathcal{O}(e^{-i\beta\lambda_L}e^{-i\beta\lambda_R})$,
respectively. From now on we denote the order of the operators as $\mathcal{O}(1,0)+\mathcal{O}(1,1)$ 
(left operators) and 
$\mathcal{O}(0,1)+\mathcal{O}(1,1)$ (right operators).\\
Eqs. (A9) can be rewritten as:
\begin{eqnarray}
\langle \hat{O}_L \rangle_{Q_L=1,Q_R=1}&=&\frac{Z_C(0,1)}{Z_C(1,1)}
[\langle \hat{O}_L \rangle^{1,0}
+\frac{Z_C(0,0)}{Z_C(0,1)}\langle \hat{O}_L \rangle^{1,1}]\nonumber\\
\langle \hat{O}_R \rangle_{Q_L=1,Q_R=1}&=&\frac{Z_C(1,0)}{Z_C(1,1)}
[\langle \hat{O}_R \rangle^{0,1}
+\frac{Z_C(0,0)}{Z_C(1,0)}\langle \hat{O}_R \rangle^{1,1}].\nonumber\\
\end{eqnarray}
The coefficients $\frac{Z_C(0,1)}{Z_C(0,0)}$ and
$\frac{Z_C(1,0)}{Z_C(0,0)}$
can be identified with
the right and left
normalization factors in the absence of inter-dot hopping, i.e,
$\frac{Z_C(0,0)}{Z_C(0,1)}\equiv \frac{1}{Z_R}$ and
$\frac{Z_C(0,0)}{Z_C(1,0)}\equiv \frac{1}{Z_L}$ and can be obtained from the left and right canonical partition
functions
of two independent single impurity problems at different chemical potentials $\mu_L$ and $\mu_R$, respectively
(see Refs. \onlinecite{NCAneq2,NCAneq1,NCAneq5}).
This way, the physical operators in the constrained ensemble can be written as:
\begin{eqnarray}
\langle \hat{O}_L \rangle_{Q_L=1,Q_R=1}
=\frac{Z_C(0,1)}{Z_C(1,1)}
[\langle \hat{O}_L \rangle^{1,0}
+\frac{1}{Z_{R}}\langle \hat{O}_L \rangle^{1,1}]\nonumber\\
\langle \hat{O}_R \rangle_{Q_L=1,Q_R=1}
=\frac{Z_C(1,0)}{Z_C(1,1)}
[\langle \hat{O}_R \rangle^{0,1}
+\frac{1}{Z_{L}}\langle \hat{O}_R \rangle^{1,1}].\nonumber\\
\end{eqnarray}
Eq. (A11) is the central result of this section.
Left (right) physical operators in the restricted ensemble
with $Q_L=1,Q_R=1$ contain two terms:\\ 
\\
{\it i)} the coefficient of the term of order $O(e^{-i\beta\lambda_{L(R)}})$
in the operator evaluated in the grand canonical ensemble plus\\
{\it ii)} the coefficient of the term of order
$O(e^{-i\beta\lambda_L}e^{-i\beta\lambda_R})$
in the operator evaluated in the grand canonical ensemble
divided by the normalization factor of the right (left)
problem without inter-dot coupling.

The new normalization factors, $\frac{Z_C(0,1)}{Z_C(1,1)}$ and $\frac{Z_C(1,0)}{Z_C(1,1)}$ can be obtained from the identities
$\langle \hat{Q}_L \rangle_{Q_L=1,Q_R=1}\equiv 1$ and $\langle \hat{Q}_R \rangle_{Q_L=1,Q_R=1}\equiv 1$.  

Now, we apply the previous projection procedure to the selfenergies of Eqs. (10) and (11).
The projection of the selfenergies can be accomplished in three basic steps.
In a first step, we follow Langreth and Nordlander in Ref. \onlinecite{NCAneq1} (see also Ref. \onlinecite{NCAneq2}): 
since the Dyson equations for $G_{L(R)}^<$ and
$B_{L(R)}^<$
contain either $G_{L(R)}^<$ or $B_{L(R)}^<$ in every term, the selfenergies that multiply these quantities
must have all terms proportional to $G_{L(R)}^<$ or $B_{L(R)}^<$ or higher projected out.
As a result we obtain from Eqs. (10) and (11) the following selfenergies ($\tau=t-t'$):
\begin{widetext}
\begin{eqnarray}
\Sigma_{L(R) , \sigma}^{r}(\tau)&=&\Bigg\{\frac{1}{N} \tilde{K}_{L(R),\sigma}^{>}(\tau)
+i(\frac{t_C}{N})^2 [\tilde{G}_{R(L),\sigma}^{r}(\tau)-\tilde{G}_{R(L),\sigma}^{a}(\tau)]
\tilde{B}_{R(L)}^{<}(-\tau)\Bigg\}\times B_{L(R)}^{r}(\tau)\nonumber\\
\Pi_{L(R)}^{r}(\tau)&=&\Bigg\{\frac{1}{N}\sum _{\sigma} \tilde{K}_{L(R),\sigma}^{<}(-\tau)
+i(\frac{t_C}{N})^2 \sum_{\sigma} [\tilde{B}_{R(L)}^{r}(\tau)-\tilde{B}_{R(L)}^{a}(\tau)]
\tilde{G}_{R(L) ,\sigma}^{<}(-\tau)\Bigg\}\times
G_{L(R) ,\sigma}^{r}(\tau).
\end{eqnarray}
Similarly,
\begin{eqnarray}
\Sigma_{L(R) , \sigma}^{<}(\tau)
&=&\Bigg\{\frac{1}{N} \tilde{K}_{L(R),\sigma}^{<}(\tau)
+i(\frac{t_C}{N})^2 \tilde{G}_{R(L),\sigma}^{<}(\tau)
[\tilde{B}_{R(L)}^{r}(-\tau)-\tilde{B}_{R(L)}^{a}(-\tau)]\Bigg\}\times B_{L(R)}^{<}(\tau)\nonumber\\
\Pi_{L(R)}^{<}(\tau)
&=&\Bigg\{\frac{1}{N}\sum _{\sigma}
\tilde{K}_{L(R),\sigma}^{>}(-\tau)
+i(\frac{t_C}{N})^2 \sum _{\sigma} \tilde{B}_{R(L)}^{<}(\tau)
[\tilde{G}_{R(L) ,\sigma}^{r}(-\tau)-\tilde{G}_{R(L) ,\sigma}^{a}(-\tau)]\Bigg\}\times
G_{L(R) ,\sigma}^{<}(\tau).
\end{eqnarray}
\end{widetext}
Where we have emphasized in our notation the structure $\Bigg\{ Kernel1+ Kernel2\Bigg\}\times propagator$.
In kernel 1,
the conduction electron propagators $\tilde{K}$ are defined in terms of the Fourier
transforms of the {\it bare} conduction electron
propagators (namely, without dot-lead coupling) as:
$\tilde{K}_{\alpha,\sigma}^{\gtrless}(\epsilon)=2\pi\sum_{k_\alpha}V^2\delta(\epsilon-\epsilon_{k_\alpha})
f_\alpha^{\gtrless}(\epsilon)$,
where $f_\alpha^{<}(\epsilon)=\frac{1}{e^{\beta(\epsilon-\mu_\alpha)}+1}$ is the Fermi function
and $f_\alpha^{>}(\epsilon)=1-f_\alpha^{<}(\epsilon)$ (see Ref. \onlinecite{NCAneq1} and below).
This way, the kernel is $\mathcal {O}(0,0)$.
The Green's functions within the kernel2 part, namely $\tilde{G}, \tilde{B}$,
{\it do not} include inter-dot hopping meaning that the kernel is $\mathcal {O}(0,1)$ for the left part and
$\mathcal {O}(1,0)$ for the right one.
This previous projection in the kernels is completely equivalent to the projection one does
in the single impurity problem:               
\begin{figure}
\includegraphics [width=0.20\textwidth,clip] {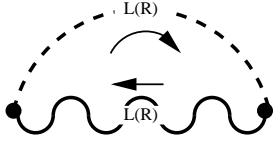}
\caption{Self-energy contribution of order $\mathcal{O}(e^{-i\beta\lambda})$ to the conduction electron 
propagator which is projected out by the constraint. The inclusion of this selfenergy contribution to the 
conduction electron propagator would give unwanted contributions of order $\mathcal{O}(1,0)$ (left side) and
$\mathcal{O}(0,1)$ (right side).}  
\label{fig17}
\end{figure}   
with the same kind of arguments one neglects terms of order
$O(e^{-i\beta\lambda})$ in the
conduction electron propagator which, in principle,
is a full propagator to be calculated in the presence of slave fermions and bosons.
Self-energy corrections to the lead electron propagators like the bubble diagram in Fig. 17 are thrown away
in the single impurity case (and also, of course, in our case).
As a consequence of this projection, one always
works with {\it bare} conduction electron propagators, which, again,
is not what one gets initially from the
unprojected NCA equations.                

In a second step we project out unwanted contributions from the propagators multiplying the kernels. 
As we mentioned previously, the left (right) kernel is of order $\mathcal {O}(0,0)$ + $\mathcal {O}(0,1)$
($\mathcal {O}(0,0)$ + $\mathcal {O}(1,0)$), meaning that the retarded propagators multiplying these kernels 
Eq. (A12)
should be of order $\mathcal {O}(0,0)$, namely {\it bare} propagators with respect to interdot. 
The corresponding lesser propagators in Eq. (A13) therefore contribute with 
$\mathcal {O}(1,0)$ (left operators) and $\mathcal {O}(0,1)$ (right operators) giving the correct order: 
$\mathcal {O}(1,0)$ + $\mathcal {O}(1,1)$ for the left operators and $\mathcal {O}(0,1)$ + $\mathcal {O}(1,1)$ 
for the right ones. 
Finally, according to Eq. (A11) the $\mathcal {O}(1,1)$ contributions should be normalized by $Z_R$ and $Z_L$
respectively.
The final set of projected selfenergies is then:
\begin{widetext}
\begin{eqnarray}
\Sigma_{L(R) , \sigma}^{r}(\tau)&=&\Bigg\{\frac{1}{N} \tilde{K}_{L(R),\sigma}^{>}(\tau)
+i(\frac{t_C}{N})^2 [\tilde{G}_{R(L),\sigma}^{r}(\tau)-\tilde{G}_{R(L),\sigma}^{a}(\tau)]
\frac{\tilde{B}_{R(L)}^{<}(-\tau)}{Z_{R(L)}}\Bigg\}\times \tilde{B}_{L(R)}^{r}(\tau)\nonumber\\
\Pi_{L(R)}^{r}(\tau)&=&\Bigg\{\frac{1}{N}\sum _{\sigma} \tilde{K}_{L(R),\sigma}^{<}(-\tau)
+i(\frac{t_C}{N})^2 \sum_{\sigma} [\tilde{B}_{R(L)}^{r}(\tau)-\tilde{B}_{R(L)}^{a}(\tau)]
\frac{\tilde{G}_{R(L) ,\sigma}^{<}(-\tau)}{Z_{R(L)}}\Bigg\}\times
\tilde{G}_{L(R) ,\sigma}^{r}(\tau).
\end{eqnarray}
\begin{eqnarray}
\Sigma_{L(R) , \sigma}^{<}(\tau)
&=&\Bigg\{\frac{1}{N} \tilde{K}_{L(R),\sigma}^{<}(\tau)
+i(\frac{t_C}{N})^2 \frac{\tilde{G}_{R(L),\sigma}^{<}(\tau)}{Z_{R(L)}}
[\tilde{B}_{R(L)}^{r}(-\tau)-\tilde{B}_{R(L)}^{a}(-\tau)]\Bigg\}\times \tilde{B}_{L(R)}^{<}(\tau)\nonumber\\
\Pi_{L(R)}^{<}(\tau)
&=&\Bigg\{\frac{1}{N}\sum _{\sigma}
\tilde{K}_{L(R),\sigma}^{>}(-\tau)
+i(\frac{t_C}{N})^2 \sum _{\sigma} \frac{\tilde{B}_{R(L)}^{<}(\tau)}{Z_{R(L)}}
[\tilde{G}_{R(L) ,\sigma}^{r}(-\tau)-\tilde{G}_{R(L) ,\sigma}^{a}(-\tau)]\Bigg\}\times
\tilde{G}_{L(R) ,\sigma}^{<}(\tau).
\end{eqnarray}
Which correspond to Eqs. (19) and (20) used in the main text. 
\end{widetext}                          


\begin{references}
\bibitem{interdot2}
The antiferromagnetic coupling due to superexchange also vanishes
in the model, because $U_L=U_R=U_{\rm intradot}\rightarrow\infty$.
\bibitem{vertex}
Vertex corrections are $\mathcal{O}(\frac{1}{N^2})$, see Ref. \onlinecite{slave3},
which makes
the NCA approximation correct to order $\mathcal{O}(\frac{1}{N})$.
\bibitem{projection}
Here, we focus on a projection procedure which is valid for diagonal (in the left/right index sense) operators. 
As we mentioned in section IVA, we can construct a fully selfconsistent NCA theory to order $\mathcal{O}(1/N)$ by 
just including diagonal propagators/self-energies.
\end{references}
\end{document}